\documentclass[a4paper,11pt]{article}
\usepackage{jcappub} 
\usepackage{natbib}
\usepackage{comment}
\usepackage[titletoc,title]{appendix}
\usepackage{graphicx}
\usepackage{amsmath}
\usepackage{breqn}
\usepackage{float}
\usepackage{subfigure}
\usepackage[usenames]{color}
\usepackage{amssymb}
\usepackage{epstopdf}
\newcommand{\bea}{\begin{aligned}}
\newcommand{\eea}{\end{aligned}}
\newcommand{\be}{\begin{equation}}
\newcommand{\ee}{\end{equation}}
\newcommand{\pr}{\partial}

\newcommand{\bse}{\begin{subequations}}
\newcommand{\ese}{\end{subequations}}

\newcommand{\rb}{\bar{r}}

\usepackage{hyperref}
\hypersetup{
     colorlinks   = true,
     citecolor    = red,
     linkcolor    = blue,
     urlcolor     = blue,
}

\renewcommand{\v}[1]{\ensuremath{\mathbf{#1}}} 
\newcommand{\bmm}{\begin{multline}}
\newcommand{\emm}{\end{multline}}

\newcommand{\mi}{\mathrm{i}}
\numberwithin{equation}{section}
\begin{document}
\title{
Extreme mass ratio inspirals in the cold vector dark matter environment
}
\author[a]{Rajesh Karmakar}
\author[b]{Debaprasad Maity}
\author[b]{Kaustubh Mukund Vispute}
\affiliation[a]{Department of Physics, Shanghai University, 99 Shangda Road, Shanghai, 200444, China}
\affiliation[b]{Department of Physics, 
Indian Institute of Technology, Guwahati, India}
\emailAdd{rajesh@shu.edu.cn}
\emailAdd{debu@iitg.ac.in}
\emailAdd{kaustubhvispute28@gmail.com}
\pagenumbering{arabic}
\renewcommand{\thesection}{\arabic{section}}

\abstract{With regard to the observed dark matter density profile in galaxies and clusters, the scalar dark matter scenario has been previously studied for potential detectability through gravitational wave observations at measurable signal-to-noise ratios. In the present study, we consider the case of dark matter described by a massive vector field, also referred to as the Proca field. The density profile in the vicinity of the black hole is explicitly constructed for a broad range of dark matter mass, $\mu\sim 10^{-10}-10^{-15}{\rm eV}$, which allows it to exhibit both particle and wave-like characteristics. While in the particle regime, the computation of the DM density distribution is analytically tractable, we find it convenient to compute the same numerically in the wave regime. Nevertheless, in the outer region, the surrounding dark matter is assumed to follow a broken power-law distribution, represented by a Navarro–Frenk–White (NFW) profile with a central spike. For the purpose of investigating the detectability of the vector dark matter in the gravitational wave spectrum, we have modelled a stellar-mass black hole ($1M_{\odot}$) inspiralling into a Schwarzschild black hole of mass $10^4M_{\odot}$ within such a vector dark matter environment. With this setup, we analyzed the dephasing in the gravitational wave strain induced by vector dark matter and performed a Fisher forecast for upcoming LISA observations, with particular emphasis on the distinctive features in both the particle and wave regimes of the dark matter. Additionally, most of the important results have been compared with the scalar dark matter case.}

\maketitle

\section{Introduction}\label{intro}

Evidence for dark matter (DM) as a dominant component of the universe is overwhelming, yet its nature remains elusive. Theoretical models broadly fall into three categories: geometric modifications of gravity (e.g. $f(R)$), particle DM from the early universe \cite{Arbey:2021gdg}, and condensates of classical fields \cite{Suarez:2013iw}. While the first two are well-studied, classical field condensates have recently attracted significant attention. A particularly interesting scenario involves black holes (BHs) surrounded by non-trivial field configurations generated through superradiance or accretion. Rotating BHs can amplify ultralight fields via superradiance, forming equilibrium field clouds, while accretion can produce purely time-dependent asymptotic configurations. Although the former is typically absent\footnote{ Note that it has been recently shown that a static BH in gravity coupled with nonlinear electrodynamics can also support the formation of superradiant bound states\cite{Dolan:2024qqr}.}, the latter phenomenon makes the static BH case particularly interesting to study. Jacobson \cite{Jacobson:1999vr} first showed that a massless scalar field with linear time dependence at infinity can persist as BH “hair,” later extended to massive scalar \cite{Clough:2019jpm, Wong:2019yoc, Ravanal:2024odh} modelling galactic DM halos \cite{Hui:2019aqm}. Recent work \cite{Hancock:2025ois} shows that the Proca field accreting onto a Schwarzschild BH can also form a Proca condensate or solitonic structure near the horizon. These studies essentially linked accretion to BH mass growth and hair formation. As far as the detectability of such hair is concerned, gravitational waves (GWs) provide a direct avenue, being inherently sensitive to gravitational effects. For this motivation, there has been an extensive effort to search for ultralight dark matter in the current and future GW observations. Particularly, the scalar case has been extensively studied \cite{Kadota:2023wlm,kim2023adiabatically}. Recently, the detectability of a coherently oscillating ultralight vector field, representing vector dark matter, has been investigated \cite{Chase:2025wwj} in the setting of upcoming LISA GW observations \cite{amaro2017laser,danzmann1996lisa,Hughes:2001ch}. In the same spirit, it is crucial to extend these analyses to realistic dark matter scenarios that align with observational constraints, spanning both the wave and particle regimes over a wide range of DM mass, and study their GW signatures.

Most galaxies are known to harbour a central BH, whose presence profoundly influences their structure and evolution \cite{Falcke:2013ola}. Surrounding the BHs, the presence of DM manifests through the rotation curve of the galaxies \cite{Battaner:2000ef}. Importantly, measurements of orbital evolution of individual stars in the galaxy suggest a nearly uniform distribution of DM surrounding the galaxy core \cite{Shen:2023kkm, GRAVITY:2020gka, Lacroix:2018zmg, Balaji:2023hmy}. On the theoretical side, cosmological N-body simulation provides the density profile, described by the the Navarro–Frenk–White (NFW) model, to fall as $\rho\sim r^{-1}$ near the BH, sometimes referred to as the cusp inner profile. Within such a cuspy DM halo, it was first shown in \cite{Gondolo_1999} that the adiabatic growth of a BH, due to baryonic accretion, contracts the orbits of nearby DM particles, leading to the formation of a spike in the density profile. For the wave DM case, the appearance of similar spikes has been investigated in Ref.\cite{kim2023adiabatically}. Such overdensity should have a luminous outcome due to DM annihilation \cite{Bertone:2002je, Merritt:2002vj}. However, the lack of observational evidence for such emission \cite{HESS:2016mib}, and the upper bound on the DM-neutrino annihilation radiation \cite{Arguelles:2019ouk, Beacom:2006tt}, favour the uniform profile.
Furthermore, computations of scattering of DM by nearby stars surrounding the galactic centre \cite{Bertone:2005hw, Sadeghian:2013laa, Ferrer:2017xwm} and formation of a nearby BH \cite{ Ullio:2001fb} suggest weakening of the spike. In this regard, formation of solitonic structure as the core profile, with the successive spike and NFW model, provides for the best-suited scenario so far. Together with the central BH, such a DM medium should have a significant effect on the dynamical evolution of a secondary compact object, such as a companion BH. Then, the impact can be further analyzed with the emitted GW from this binary BH in the presence of DM.  

It is worth mentioning that, in studying the impact of DM on GW in the context of binary BH, several works have followed an agnostic approach towards DM, i.e., remaining independent of the specific characteristics of DM, whether scalar or vector \cite{Wade:2025rkk, Mitra:2025tag, Nichols:2023ufs, Eda:2013gg}. On the other hand, such environment effects have also been included in the spacetime background itself in studying the motion of the inspiral \cite{Rahman:2023sof}. Building on these studies, we investigate whether it is possible to probe the fundamental nature of DM in our analysis. Setting aside the various dynamical processes \cite{Bertone:2005hw, Sadeghian:2013laa, Ferrer:2017xwm, Ullio:2001fb} that may significantly affect the DM density distribution, examining extreme mass ratio inspirals (EMRIs) embedded in a vector DM medium spanning a wide range of masses presents an interesting extension to studies focused on scalar DM. For this purpose, we focus on a stellar-mass BH inspiralling into an intermediate-mass black hole embedded in a vector DM halo represented by the Proca field. For such a scenario, the stellar mass BH can also acquire vector charge and thereby influence the GW waveform\footnote{For analysis in EMRI with the stellar mass companion carrying scalar charge see Ref.\cite{Zi:2025lio, Barsanti:2022vvl}.} \cite{Zhang:2022hbt, Zi:2024lmt, Zhang:2024ogc, zi2025eccentric}. However, in the present analysis, we neglect such an effect and focus solely on the effects of dynamical friction from the surrounding vector DM on the orbital dynamics \cite{Fell:2023mtf}. This also affects the backreaction due to GW emission, whereas the drag force due to DM accretion turns out to be subleading. To construct the DM density profile, we find the solution of the Proca field in the background of a central BH described by the Schwarzschild spacetime and compute the density profile normalized at a finite distance away from the horizon with a spike profile, subsequently joined with the NFW model. This distance scale is determined by the self-gravity of the DM \cite{Hui:2019aqm}. A similar analysis with an average density profile has been studied previously \cite{Hancock:2025ois}. However, the analysis of GW and the detectability of the presence of such a vector DM profile have not yet been explored. With this motivation, we have computed the GW strain amplitude due to a stellar mass BH inspiralling the central BH embedded in such a vector DM environment. The impact of the DM medium on the dephasing of the GW has been analyzed. The consideration of EMRI makes the frequency of the GW to lie within the detectable range of the upcoming LISA observation \cite{amaro2017laser}. Given this opportunity, utilizing the LISA noise sensitivity, we have performed the Fisher forecasting of the important parameters in our model.

The rest of the paper has been organized as follows. In Sec.\ref{sec.field.eqn.soln}, we presented the governing equations of the Proca field describing the vector DM and expressed the solutions in terms of the confluent Heun function. These solutions, further analyzed, in Sec.\ref{sec.particle.wave.soln}, for particle and wave regimes. This section also serves as an introduction to the self-gravity radius due to the presence of DM. Next, in Sec.\ref{sec.model.density}, we have discussed the NFW DM density model with the associated spike profile. This model is then utilized in Sec.\ref{sec.normalization} to normalize the core density profile within the self-gravity radius. Sec.\ref{sec.bbh.forces} introduces the quasi-circular motion of a stellar mass BH around the central BH, and the impact on the orbital evolution is investigated by computing the forces for gravitational backreaction, dynamical friction and accretion. In Sec.\ref{sec.dephasing}, we have analyzed the dephasing of GW, emitted by the EMRI, due to the vector DM. Sec.\ref{sec.fisher} is dedicated to the Fisher analysis for the GW with the LISA. We have concluded with future outlook in Sec.\ref{sec.conclusion}.
\section{Proca field in the background of the Schwarzschild spacetime}\label{sec.field.eqn.soln}
To investigate the properties of vector DM in a static, spherically symmetric background, we begin with the approximation that the field does not backreact on the BH geometry, at least up to a certain radius (which will be clarified in the next section). Within this radius, the spacetime is well described by the Schwarzschild metric and the corresponding line element is given by
\be
ds^2=-f(r)dt^2+\frac{1}{f(r)}dr^2+r^2d\theta^2+r^2\sin^2\theta d\varphi^2,
\ee
with $f(r)=1-r_h/r$, where $r_h$ stands for the Schwarzschild radius. For BH mass $M$, the event horizon radius appears to be $r_h=2M$. On this spacetime, we consider the following Lagrangian density for a minimally coupled massive vector boson,
\be
    \mathcal{L} = -\frac{1}{4}F_{\alpha\beta}F^{\alpha\beta} - \frac{1}{2}\mu^{2}A_{\alpha}A^{\alpha},
\ee
where $F_{\alpha\beta} = \nabla_{\alpha}A_{\beta} - \nabla_{\beta}A_{\alpha}$ is the field-strength tensor, and $\mu$ denotes the mass of the vector field, physically representing DM mass for the present context. Whereas, the $ \nabla_\mu$ stands for the covariant derivative in curved spacetime. The above Lagrangian provides for the governing equation of motion of the Proca field,
 \begin{equation}\label{eq.eom1}
    \nabla_\mu F^{\mu\nu} + \mu^{2}A^{\nu} = 0.
 \end{equation}
 The presence of a mass term automatically leads to the Lorentz condition, $\nabla_\mu A^\mu=0$.
 Nevertheless, given the spherically symmetric nature of the Schwarzschild spacetime, with the minimal coupling of the vector field, it is convenient to decompose the field components utilizing  the vector spherical harmonics in the following manner,
\be\label{decom.em}
\bea
&A_t(t,\v r)=\sum_{lm} b^{lm}(t,r)Y_{lm}(\theta,\varphi) ,\\
&A_r(t,\v r)=\sum_{lm}h^{lm}(t,r)Y_{lm}(\theta,\varphi) ,\\
&A_{s}(t,\v r)=\sum_{lm}\left[k_{lm}(t,r)\Psi^{lm}_{s}(\theta,\varphi)+a_{lm}(t,r)\Phi^{lm}_{s}(\theta,\varphi)\right],
\eea
\ee
where $\Psi^{lm}_{s}(\theta,\varphi)$ and $\Phi^{lm}_{s}(\theta,\varphi)$ represent the vector spherical harmonics, constructed out of scalar spherical harmonics $Y^{lm}$ \cite{Barrera}. Note that the symbol $s$, here, denotes the $\theta$ and $\varphi$ components. Substituting the decomposed field components in the governing equation \eqref{eq.eom1}, we derive the following two coupled differential equations \cite{konoplya2006massive},
\be\label{t.eq}
f(r)\pr_r(r^2(\pr_t h^{lm}-\pr_r b^{lm}))+l(l+1)(b^{lm}-\pr_t k^{lm})+\mu^2r^2 b^{lm}=0,
\ee
\be\label{r.eq}
r^2f(r)^{-1}\pr_t(\pr_t h^{lm}-\pr_r b^{lm})+l(l+1)(h^{lm}-\pr_r k^{lm})+\mu^2r^2 h^{lm}=0.
\ee
The above field components belong to the even parity sector. To simplify the analysis further in the rest of our discussion,  we consider only the even parity s-wave, $l=0$, monopole mode. Recently, it has been shown that dipole Proca modes also have a very similar density profile \cite{Hancock:2025ois}. However, considering dipole modes, we find no significant or interesting changes to our conclusions in the mass range under consideration, $\mu\gtrsim 10^{-15}{\rm eV}$. For even higher modes, as the centrifugal barrier becomes dominant, the density profile is expected to be subdominant \cite{Yavetz:2021pbc} as compared to the initial modes, in the region of our interest. On the other hand, as far as the comparison with previous work on scalar DM \cite{Kadota:2023wlm, kim2023adiabatically} is concerned, and since all the effects and observables we consider are gravitational and essentially depend on the DM density distribution, we find the monopole mode sufficient.  Nevertheless,  the analysis for lower DM mass, $\mu\lesssim 10^{-15}{\rm eV}$, including the higher modes, is left for future investigation. Now, considering the monopole mode, we perform the following operation, $\pr_r\eqref{t.eq}-\pr_t\eqref{r.eq}$. Then introducing a gauge invariant variable, $\chi(t,r)=r(\pr_t h^{00}-\pr_r b^{00})$, we obtain the master equation
\be
\pr^2_{r_*}\chi-\pr^2_t\chi-\left(1-\frac{r_h}{r}\right)\left\{\frac{2}{r^2}-\frac{3r_h}{r^3}+\mu^2\right\}\chi=0,
\ee
where $r_*$ represents the tortoise coordinate, defined as $dr_*=dr/f(r)$. As a natural consequence of a static spacetime background, having a time-like Killing vector field, $\pr_t$, allows separation of the temporal part of the field. Hence, we substitute $\chi(t,r)=e^{-\mi \omega t}\bar{\chi}(r)$, where $\omega$ denotes the frequency of the field, and the above equation yields
\be\label{rad.chibar.eqn}
\frac{d^2 \bar{\chi}}{dr^2_*}+\left[\omega^2-\left(1-\frac{r_h}{r}\right)\left\{\frac{2}{r^2}-\frac{3r_h}{r^3}+\mu^2\right\}\right]\bar{\chi}=0.
\ee
The general solution of this equation can be expressed in terms of the confluent Heun (HeunC) function as \cite{Bezerra:2013iha} (see Appendix.\ref{Heun_derive} for the detailed derivation),
\be
\bea
\bar{\chi}&= c_{1}e^{ikr}r^{3}(r-r_h)^{i\omega r_h}\text{HeunC}\left(-2ik r_h,2i\omega r_h,4,-k^2r^2_h - \omega^2r^2_h,k^2r^2_h+\omega^2r^2_h+2,1-\frac{r}{r_h}\right)\\
&+c_{2} e^{-ikr}r^{3}(r-r_h)^{-i\omega r_h}\text{HeunC}\left(2ik r_h,-2i\omega r_h,4,-k^2 r^2_h - \omega^2 r^2_h,k^2r^2_h+\omega^2r^2_h+2,1-\frac{r}{r_h}\right),
\eea
\ee
where $k=\sqrt{\omega^2-\mu^2}$. In the near horizon limit, $r\rightarrow r_h$,  by definition $\text{HeunC} \approx 1$ \cite{Bezerra:2013iha, Vieira:2021nha}. Consequently, the above solution, with the temporal part, in this limit reads,
\be\label{soln.nearhorizon}
    \chi(t,r) \approx c_{1}e^{-i\omega(t-r_{*})} e^{-i(\omega -k)} + c_{2} e^{-i\omega(t+r_{*})}e^{i(\omega - k)}.    
\ee
This helps us to identify the ingoing and outgoing modes of the solution. Imposing the ingoing boundary conditions near the BH horizon, as required by causality, fixes one of the constants as $c_{1} = 0$. Conformity with this imposition forces the solution to be
\be\label{soln.in.part}
    \bar{\chi}\approx c_{2}r^3 e^{-ikr}(r-r_h)^{-i\omega r_h}\text{HeunC}\left(2ikr_h,-2i\omega r_h,4,-k^2r^2_h- \omega^2 r^2_h,k^2r^2_h+\omega^2r^2_h+2,1-\frac{r}{r_h}\right).
\ee
For the Proca solution above, describing collisionless non-relativistic cold DM on galactic scales, the dynamics are dominated by the mass term of the field \cite{Armendariz-Picon:2013jej, Kitajima:2023fun}. As a result, the field oscillates at a frequency $\omega\sim \mu$, and, according to our previous definitions, this corresponds to $k\sim 0$, reflecting the slowly varying, nearly uniform nature of the cold DM distribution. With this consideration, the above expression involving the confluent Heun function can be numerically evaluated. Moreover, in the presence of a DM environment, it would be natural to utilize some well-motivated models that are compatible with the observations to determine the overall constant factor. However, this normalization must be performed at a specific areal radius, which we define in the next section, where we also discuss the dark matter mass regimes, which determine whether it behaves predominantly as a wave or as particles.

\section{Particle-wave nature of the vector dark matter and its self-gravity}\label{sec.particle.wave.soln}
In the preceding analysis, we obtained the solution of the DM field, assuming that the spacetime geometry is primarily governed by the central BH. However, this is only true near the BH, up to a certain distance, within which the BH mass dominates over the enclosed mass of DM  \cite{Hui:2019aqm, Ravanal:2024odh}. In this region, for such a self-gravitating system, the average gravitational potential energy will be dominated by the BH, and can be expressed as
\be
\langle V\rangle=-\frac{GMM_{\rm sg}}{r_{\rm sg}},
\ee
where $r_{\rm sg}$ is the radius within which the self-gravity of the DM can be neglected. Whereas, $M_{\rm sg}$ is the total mass of the DM enclosed within $r_{\rm sg}$. Henceforth, we will refer to $r_{\rm sg}$ as the self-gravity radius. Getting back, the average kinetic energy for the BH-DM system can be inferred, in the rest frame of the BH as
\be
\langle T\rangle=\frac{1}{2}M_{\rm sg}\langle v^2\rangle.
\ee
Now, the virial theorem for the self-gravitating system states that
\be
2\langle T\rangle+\langle V\rangle=0.
\ee
Applying it to the BH-DM system, this translates to the following formula for the velocity dispersion,
\be
\langle v^2\rangle=\frac{2GM}{r_{\rm sg}}\equiv \frac{r_h}{r_{\rm sg}}.
\ee
Alternatively, the self-gravity radius, $r_{\rm sg}$, can be evaluated knowing the velocity dispersion of the DM. Such as, in galactic DM halo, with the typical velocity $ \sim 10^2 {\rm km/s}$ \cite{Sales:2007hp, Battaglia:2005rj}, the above equation suggests that $r_{\rm sg}\sim 10^6 r_h$ \cite{Hui:2019aqm}. It is important to note that the observed velocity dispersion is always associated with a specific spatial scale, as it is locally measured and not restricted to the BH's sphere of influence. However, the dispersion typically increases towards the smaller radius \cite{Sales:2007hp}. Therefore, it is reasonable to adopt a higher velocity dispersion in our context. Accordingly, we consider $v\sim 10^3{\rm km/s}$, which corresponds to a self-gravity radius, $r_{\rm sg}\sim 10^5r_h$ \cite{Hui:2019aqm}. Nevertheless, in the exterior to this surface, $r>r_{\rm sg}$, the backreaction may not be negligible, and the Schwarzschild geometry might be modified. This motivates us to analyse solutions only within the self-gravity radius and to normalize them against existing models at $r\sim r_{\rm sg}$.

Importantly, the velocity dispersion $v\sim 10^3{\rm km/s}$ can be utilized to distinguish the particle and wave regimes of the DM. The particle regime is determined by $\mu v r>1$, i.e. the de-Broglie wavelength is much smaller than the characteristic scale, $r$. In the presence of the BH, it is convenient to define this regime in the scale of horizon radius as $\mu vr_h>1$. In the present analysis, we will consider the mass of the central BH to be $10^4M_\odot$, for which the particle regime translates to
\be
\mu\gtrsim 10^{-12}\left(\frac{10^4 M_{\odot}}{M}\right)~{\rm eV}.
\ee
It is worth noting that whenever $\mu vr_h>1$, the condition $\mu v r>1$ also holds. Hence, considering heavier mass, for example, $\mu=10^{-10} {\rm eV}$, as indicated by the above bound, guarantees the particle regime across the entire spatial region exterior to the horizon. In the complementary regime $\mu vr_h<1$, we have considered $10^{-14} {\rm eV}$ and $10^{-15}{\rm eV}$. In this regime, the DM field exhibits interesting characteristics. For example, with $\mu \sim 10^{-14}$, the DM field will behave as waves for $r\lesssim 10^2r_h$ and in the exterior as particles. 

In both the particle and wave regimes, the solutions, expressed in terms of confluent Heun function \eqref{soln.in.part}, can be numerically evaluated. However, it is interesting to explore whether the analytical solution can be obtained in certain limits. For this purpose, we solve \eqref{rad.chibar.eqn} in the asymptotic limit (the details have been given in the Appendix \ref{append.asymp.soln}), and the solution takes the following form, 
\be\label{soln.rinfty}
\bea
\chi(t,r)=c_3 e^{-\mi \mu t}r^{1/4}e^{2\mi \mu \sqrt{rr_h}}\left(1+\mathcal{O}\left(\frac{1}{\mu\sqrt{rr_h}}\right)\right)+ c_4 e^{-\mi \mu t}r^{1/4}e^{-2\mi \mu \sqrt{rr_h}}\left(1+\mathcal{O}\left(\frac{1}{\mu\sqrt{rr_h}}\right)\right).
\eea
\ee
Naively, it is understood that the above solution is valid as far as $r>1(/\mu^2r_h)$.  However, in the heavy mass regime, $\mu r_h>1$, the effective potential is dominated by the Newtonian potential term, $-\mu^2r_h/r$ \eqref{rad.chibar.eqn}, i.e., the potential barrier disappears. Consequently, the above solution holds throughout the domain, extending up to the near-horizon region. Importantly, the solution \eqref{soln.rinfty} should also satisfy the ingoing condition near the horizon. As a result, the solution in this regime, $\mu r_h>1$, reads as
\be\label{soln.chi.heavymass}
    \chi(r) =c_4 r^{1/4} e^{-2i\mu\sqrt{rr_h}}.
\ee
Notably, in the heavy DM mass regime, considering $\mu  r_h\geq 1$, implies the following lower bound on the DM mass, $\mu\gtrsim 6.7\times 10^{-15}(10^4 M_{\odot}/M)~{\rm eV}$. Therefore, in line with our previous analysis, the above solution appropriately captures the particle-like dark matter regime, and also encompasses the case, such as for $\mu = 10^{-14}{\rm eV}$, where the dark matter exhibits both particle- and wave-like behaviour. 

\begin{figure}
    \centering
\includegraphics[scale=0.5]{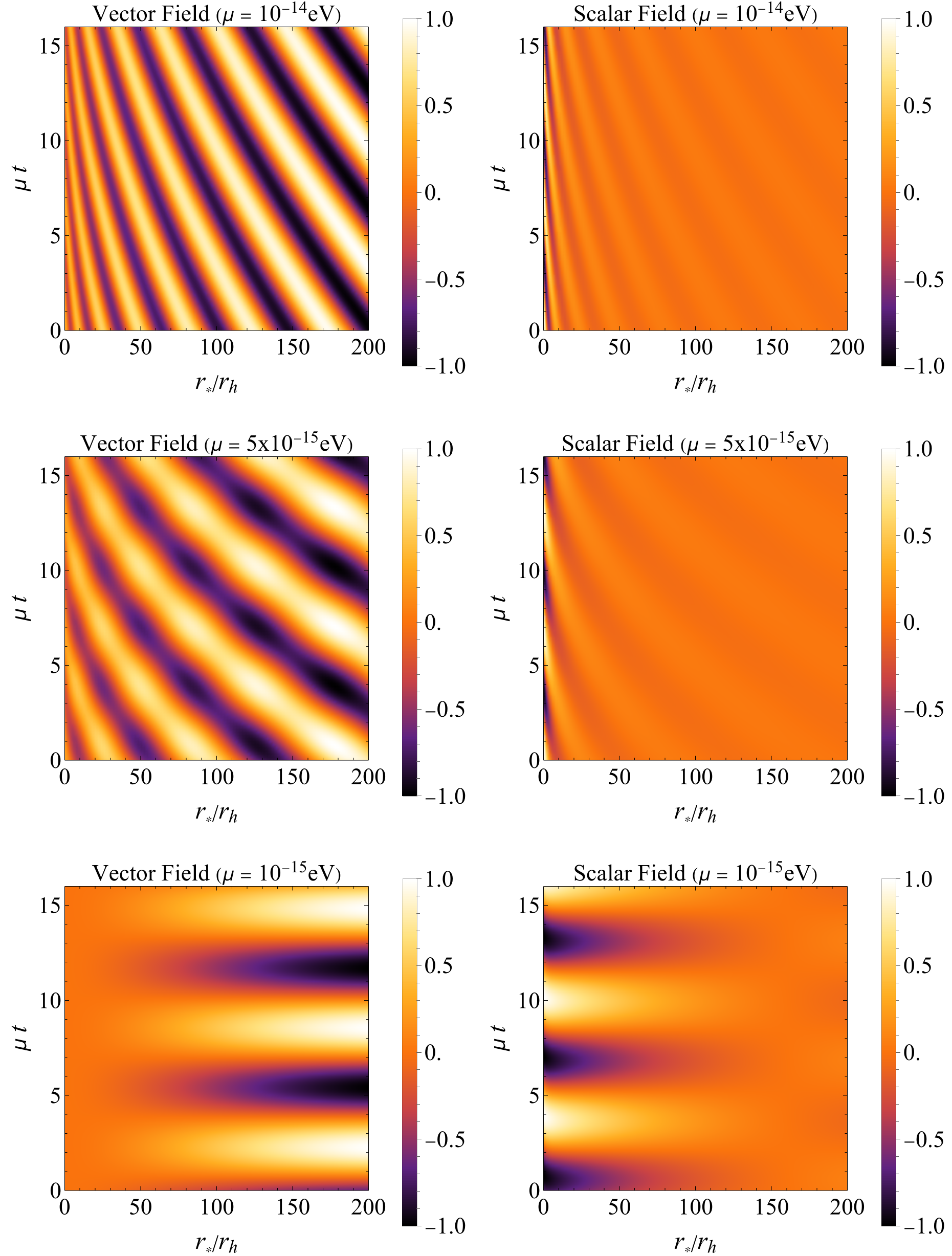}
\caption{Vector and scalar field profile as function of $t$ and $r_*$ for three values of DM mass, $\mu = 10^{-14}\rm eV,~5\times10^{-15}\rm eV,~ 10^{-15}\rm eV$ with the central BH mass $M_{\rm BH} = 10^4M_{\odot}$. Each field profile is scaled by its respective maximum value over the specified ($t,r_*$) domain.}\label{fig:numericalsoln}
\end{figure}

In the low mass regime, $\mu r_h\lesssim 1$, such as for $\mu=10^{-15} {\rm eV}$, although the confluent Heun function \eqref{soln.in.part}  can be appropriately expanded, we find it difficult to extend the solution for finite or large radial distances with various scales as discussed for the scalar case \cite{Hui:2019aqm}. Therefore, in this case, we have numerically evaluated the solution. In doing so, one will be left with the overall unknown constant, which should be fixed using an existing density profile model, such as NFW with a spike, as we discuss in the following sections. However, for visual realization of the numerical solution, we have illustrated the real part of the solution in Fig.\ref{fig:numericalsoln} by scaling with the maximum value in the ($\mu t, r_*$) domain. In the top panel of Fig.~\ref{fig:numericalsoln}, for heavier mass, $\mu=10^{-14}{\rm eV}$, the density plot shows closely spaced, diagonal wavefronts, indicative of the high temporal frequency $\omega \simeq \mu$ and the correspondingly short de Broglie wavelength, which allows radial accumulation near the BH. As the mass decreases, the field develops broader spatial fringes and slower temporal variation, as has been presented in the lower panels of the same figure. For the lowest-mass under consideration, $\mu=10^{-15}{\rm eV}$, illustrated in the last panel of this figure, the de Broglie wavelength becomes sufficiently large that the field effectively behaves as a single coherent mode, and the spatial pattern correspondingly becomes smoother. Before moving forward, we emphasize once more that, in what follows, the particle and wave regimes are referred to relative to the de Broglie wavelength of the DM, as discussed above, while the heavy and low mass regimes are referred to by comparing the DM Compton wavelength to the size of the BH horizon \cite{Traykova:2023qyv}.
\section{NFW density profile with the spike model}\label{sec.model.density}
The Navarro–Frenk–White (NFW) profile is a widely used model for describing the density structure of DM halos, considering that the universe is governed by $\Lambda$CDM cosmology. It captures the universal behaviour of halos across a wide range of masses, featuring a steeply rising density toward the centre and a gradually declining density at larger radii. Its simplicity and consistency with numerical simulations make the NFW profile a standard tool for studies of galaxy formation, halo dynamics, and the distribution of dark matter in the universe. In this model, the density profile \cite{Navarro_1996,Navarro_1997} is given by,
\be
\rho_{\rm NFW}(r)=\frac{\rho_{\rm crit}\delta_{\mathfrak{c}}}{(r/r_{sc})(1+r/r_{sc})^2},
\ee
where $\rho_{\rm crit}$ is the critical density of the universe, given by $\rho_{\rm crit} = 2.775 h^2 \times 10^{11}M_{\odot}$, with $h$ representing the reduced Hubble's constant. Whereas, $r_{sc}$ represents the characteristic radius, quantified as $r_{sc}=r_{200}/\mathfrak{c}$ with $r_{200}$ determining the mass of the BH as $M_{200}=200(4/3)\pi r^3_{200}\rho_{\rm crit}$. On the other hand, the DM concentration parameter $\mathfrak{c}$ can be obtained from the $\mathfrak{c}-M$ relation, which applies to all halo masses for the low-redshift regime $(z \ll 4)$ \cite{Correa_2015},
\begin{figure}
    \centering
\includegraphics[scale=0.38]{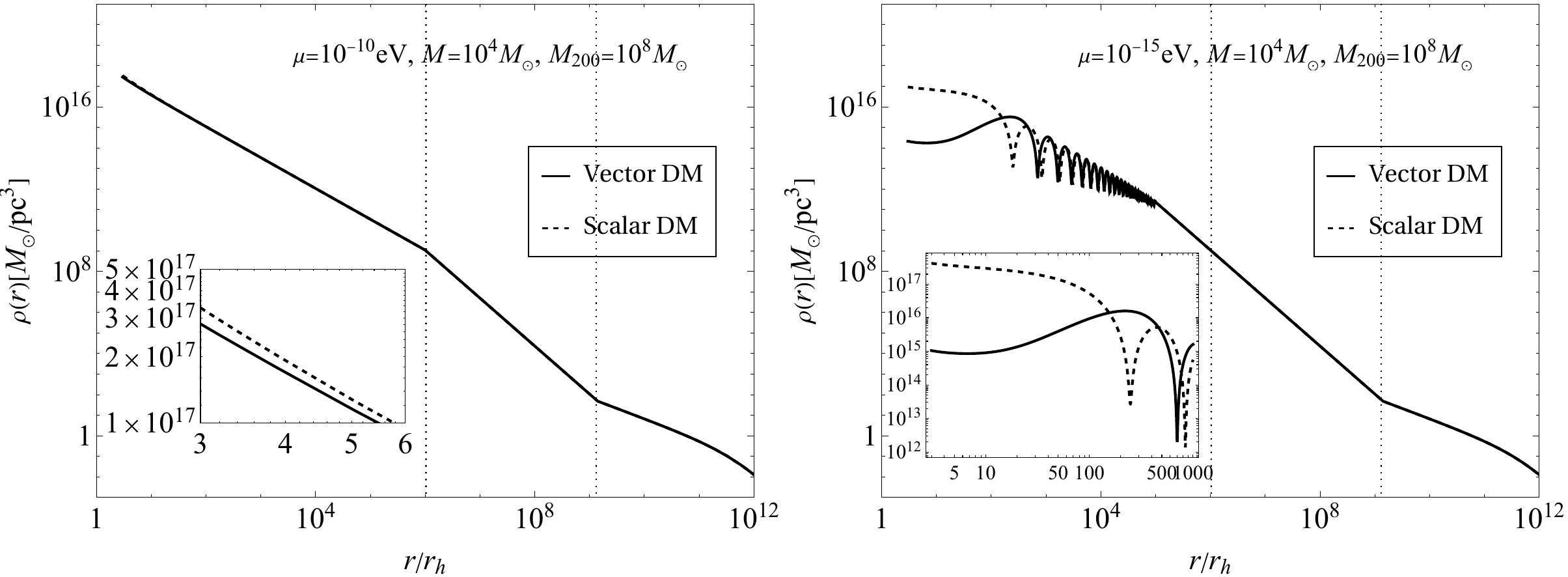}
\caption{Vector and scalar DM density profile, normalized with the NFW-spike Model, around the central BH of mass $M_{\rm BH} = 10^4M_{\odot}$, fixing $M_{200} = 10^8 M_{\odot}$, and DM field mass in the {\bf left panel:} $\mu =10^{-10}\rm eV$, and in the {\bf right panel:} $\mu =10^{-15}\rm eV$. The consecutive vertical dotted lines represent the position of $r_{\rm sg}$ and $r_{sp}$, respectively (see \eqref{eq:wholedensity}). The inset figure in the left panel shows an enlarged view of the magnitude of the density profile near the BH. Whereas, in the right panel, for the low mass regime, only a few initial oscillations have been shown with the application of the smearing procedure (see \eqref{eq:smearing}). In this panel, the inset figure depicts the enlarged version of the density profiles near the BH.}\label{fig:rhovec1}
\end{figure}
\be
\begin{aligned}
\log _{10}\mathfrak{c} & =\alpha_0+\alpha_1\log _{10}\left(M / \mathrm{M}_{\odot}\right)\left[1+\alpha_2\left(\log _{10} M / \mathrm{M}_{\odot}\right)^{2}\right], \\
\alpha_0 & =1.62774-0.2458(1+z)+0.01716(1+z)^{2}, \\
\alpha_1 & =1.66079+0.00359(1+z)-1.6901(1+z)^{0.00417}, \\
\alpha_2 & =-0.02049+0.0253(1+z)^{-0.1044} .
\end{aligned}
\ee
The other unknown parameter $\delta_{\mathfrak{c}}$, usually referred to as characteristic overdensity, is related to $\mathfrak{c}$ in such a way that the mean density within $r_{200}$ becomes $200\rho_{\rm crit}$, and can be expressed as, 
\be
\delta_{\mathfrak{c}}=\frac{200}{3}\frac{\mathfrak{c}^3}{\ln(1+\mathfrak{c})-\mathfrak{c}/(1+\mathfrak{c})}.
\ee

While the above formulas, constituting the NFW model, suitably describe the DM halo density in the outermost region (to be defined shortly) of a galaxy, they deviate due to the presence of a central supermassive BH. Specifically, the orbits of dark matter particles are compressed inward, which leads to the formation of a spike in the dark matter distribution near the centre of the galaxy\cite{Gondolo_1999, Sadeghian_2013}. Specifically, it is shown that, in the outermost region, if the initial halo follows a power law cusp $\rho_{i}(r) \propto r^{-\gamma_i}$, the resulting spike profile towards the core region becomes,
\be
\rho_{\rm spike}(r) = \rho_{\rm sp}\left(\frac{r}{r_{\rm sp}}\right)^{-\gamma_{\rm sp}}, \hspace{0.3cm} \text{with}\hspace{0.2cm} \gamma_{\rm sp} = \frac{9-2\gamma_i}{4-\gamma_i},
\ee
where $\rho_{\rm sp}$ is the density at $r=r_{\rm sp}$ and $\gamma_{\rm sp}$ is the slope of the spike. Setting the initial DM profile to be NFW beyond $r_{\rm sp}$, as described earlier, implies $\gamma_i=1$, yielding $\gamma_{\rm sp}=7/3$. The continuity between the spike and NFW regions is enforced at $r_{\rm sp}$ through the condition
\be\label{eq:continuity}
\rho_{\rm NFW}(r=r_{\rm sp}) = \rho_{\rm spike}(r=r_{\rm sp}).
\ee
On the other hand, conventionally, the total DM mass within $5r_{\rm sp}$ is considered to be twice the BH mass \cite{Eda:2014kra, Merritt:2003qk},
\be\label{eq:bhdmmass}
\int_{r_{\rm min}}^{5 r_{\rm sp}} \rho_{DM}(r) 4\pi r^2 dr = 2M.
\ee
The unknown parameters, $\rho_{\rm sp}$ and $r_{\rm sp}$ are determined by the previous two conditions \eqref{eq:continuity} and \eqref{eq:bhdmmass}. Whereas, in principle, $r_{\rm min}$ could be taken all the way down to $r=r_h$. However, earlier studies of collisionless DM suggest that for $r\lesssim 3r_h$, i.e., approximately within the innermost stable circular orbit (ISCO), the orbits become unstable and the DM is no longer gravitationally bound, causing the density to drop to zero \cite{Sadeghian_2013, Gondolo_1999}. Consequently, we impose a cutoff in the profile by setting $r_{\rm min}=3r_h$. Importantly, we adopt this particle-based spike profile also for the low mass regime DM. As discussed in Sec.\ref{sec.particle.wave.soln}, for the mass range $\mu\sim 10^{-14}-10^{-15} {\rm eV}$ the DM field behaves as a wave only within $10^2-10^3r_h$ and is effectively particle-like in the exterior. Hence, beyond $r_{\rm sg}\sim 10^5r_h$, a similar spike profile remains consistent (see Ref. \cite{kim2023adiabatically} for a related discussion). In summary, we first construct the Spike-NFW model for the DM regardless of its nature as per the method described above, then use it beyond $r_{\rm sg}$ for both the particle and wave regimes\footnote{A similar line of argument is employed in \cite{Kadota:2023wlm}, where, for the case of self-interacting (collisional) scalar DM, the collisionless spike model is adopted to describe the DM density outside the core region. For a GW analysis considering DM density modelled solely by a spike profile, see \cite{Coogan:2021uqv}.}. On the other hand, it is well known that DM can form a solitonic core or nearly uniform density profile \cite{Shen:2023kkm, GRAVITY:2020gka, Lacroix:2018zmg, Balaji:2023hmy} in the innermost region of a galaxy. Hence, it is imperative to go beyond the spike model and construct the density profile with the solution of the Vector DM interacting with the BH background. For this purpose, we normalize the solution of the Proca field, which we have obtained and discussed in the previous section, in a way that ensures its compatibility with the outer density profile in the following manner:
\be
    \rho(r)=\left\{\begin{array}{ll}
\rho_{\rm sg}(r), & r_{\min } \leq r<r_{\rm sg},\\
\rho_{\mathrm{spike}}(r), & r_{\rm sg } \leq r<r_{\rm sp},\\
\rho_{\mathrm{NFW}}(r), & r_{\mathrm{sp}} \leq r.
\end{array}\right.\label{eq:wholedensity}
\ee
We should clarify again that $\rho_{\rm sg}$ stands for the density profile as per our analysis with the vector DM field in the interior of the self-gravity radius ($r_{\rm sg}$). In the following section, we will discuss the procedure of normalizing the Proca solution utilizing the above NFW density profile with the spike model near the self-gravity radius.
\section{Normalization of the density of the Proca fields}\label{sec.normalization}
We normalize the field in such a way that the energy density will satisfy its estimated value in some galactic or extragalactic medium. For this, we first evaluate the energy density of the field at spatial infinity, and match existing models on the density of the vector DM particles beyond the self-gravity radius. The density of the vector DM can be expressed (see Appendix \ref{append.proca.density}) in the following manner in terms of the Proca field,
\be\label{eq.energymomentumtt}
\rho_{\rm sg}(r)=-{T^t}_t =\frac{1}{4\pi f(r)}\left[\frac{\chi^2 f(r)}{2r^2} +\frac{(\partial_t \chi)^2 }{2\mu^2r^2} +\frac{f^2(r)}{2\mu^2 r^4}\left[\partial_r(r\chi)\right]^2\right].
\ee
Substituting $\chi(t,r) = \bar{\chi}(r)e^{-i\mu t}$, the above expression simplifies to
\be
\rho_{\rm sg}(r)=\frac{|\bar{\chi}(r)|^2}{4\pi r^2f(r)}\left[\frac{1}{2}\left(1+f(r)\right)+\frac{f^2(r)}{2\mu^2 r^2}\right].
\ee
To simplify the subsequent computations, we assume in the above expression that the radial gradient of the Proca field is subleading. This assumption is justified in the heavy-mass regime, where the effective potential is dominated by the Newtonian contribution as mentioned in Sec.\ref{sec.particle.wave.soln}. Nevertheless, we evaluate the DM density at a finite distance away from the BH, $r_{\rm sg}$, i.e., the distance at which self-gravity starts to affect the background spacetime. The above expression of the DM density, evaluated at $r=r_{\rm sg}$, reads as,
\be\label{rho_rsg}
\rho_{\rm sg}(r=r_{\rm sg})=\frac{|\bar{\chi}(r)|^2}{4\pi r^2f(r)}\left[\frac{1}{2}\left(1+f(r)\right)+\frac{f^2(r)}{2\mu^2 r^2}\right]\Bigg|_{r=r_{\rm sg}}.
\ee
{\it Normalization in the heavy dark matter mass regime:}
From the discussion of Sec.\ref{sec.particle.wave.soln}, it is understood that as far as the DM mass falls in the heavy mass range, $\mu r_h\gtrsim 1$, the analytical solution can be straightforwardly obtained. For this reason, the subsequent computation is also analytically computed. Substituting the solution \eqref{soln.chi.heavymass} for this regime in \eqref{rho_rsg}, we fix the normalization constant $c_4$ in terms of the energy density at $r_{\rm sg}$ as,
\be
c_4 = \sqrt{\frac{4\pi f(r_{\rm sg})\rho(r=r_{\rm sg})}{\left(\frac{1}{2}(1+f(r_{\rm sg}))+\frac{f^2(r_{\rm sg})}{2\mu^2 r_{\rm sg}^2}\right)}}r_{\rm sg}^{3/4}.
\ee
Bringing in the NFW Model with spike profile, as discussed in \ref{sec.model.density}, the completely determined expression for $c_4$ reads as
\be
c_4 =\sqrt{\frac{4\pi f(r_{\rm sg})\rho_{\rm sp}\left(\frac{r_{\rm sg}}{r_{\rm sp}}\right)^{-\gamma_{\rm sp}}}{\left(\frac{1}{2}(1+f(r_{\rm sg}))+\frac{f^2(r_{\rm sg})}{2\mu^2 r_{\rm sg}^2}\right)}}r_{\rm sg}^{3/4}.
\ee
With this known expression, the normalized solution in the region $r_h \le r < r_{\rm sg}$ can be expressed as,
\be
\chi(t,r) =\sqrt{\frac{4\pi f(r_{\rm sg})\rho_{\rm sp}\left(\frac{r_{\rm sg}}{r_{\rm sp}}\right)^{-\gamma_{\rm sp}}}{\left(\frac{1}{2}(1+f(r_{\rm sg}))+\frac{f^2(r_{\rm sg})}{2\mu^2 r_{\rm sg}^2}\right)}}r^{1/4} e^{-i\mu t}e^{-2i\mu\sqrt{r r_h}},
\ee
Once the solution is fully determined, the next step is to substitute it into the full expression of the energy-momentum tensor \eqref{eq.energymomentumtt} to evaluate the density profile. In doing so, we obtain
\be
\rho_{\rm sg}(r)=\frac{f(r_{\rm sg})\rho_{\rm sp}\left(\frac{r_{\rm sg}}{r_{\rm sp}}\right)^{-\gamma_{\rm sp}}}{\left(\frac{1}{2}(1+f(r_{\rm sg}))+\frac{f^2(r_{\rm sg})}{2\mu^2 r_{\rm sg}^2}\right)} r_{\rm sg}^{3/2}\frac{r^{1/2}}{f(r)}\left[\frac{f(r)}{2r^2} +\frac{1}{2r^2} +\frac{f^2(r)}{32\mu^2 r^4}(25+16 rr_h\mu^2)\right].
\ee
The density profile for the vector DM in the heavy mass regime has been plotted in the left panel of Fig.\ref{fig:rhovec1}. For comparison, we have also included the density profile for scalar DM with monopole mode (see Appendix \ref{append.scalar.density} for mathematical details). For a large DM mass (particle regime), we see that the density steeply rises near the BH, given that the potential barrier almost vanishes, and large transmission (see also the discussion given in Appendix \ref{append.transmission}), although with a lower slope than the spike for both the scalar and vector DM in the same manner. However, the vector DM density is slightly lower than the scalar case, as in the heavy mass regime, although the mass term dominates the effective potential, a small difference in the structure of the stress-energy tensor could be the reason for this slight disparity between the vector and scalar DM.

{\it Normalization in the low dark matter mass regime:} As stated in Sec.\ref{sec.particle.wave.soln}, the Proca solution in the low-mass regime ($\mu r_h < 1$) was obtained numerically. We now outline this procedure, following the same steps as in the heavy-mass case. First, the DM density is obtained by substituting the numerically evaluated solution, Eq.~\eqref{soln.in.part}, into the density component of the energy–momentum tensor, Eq.~\eqref{eq.energymomentumtt}. Notably, the spatial variation of the energy–momentum tensor is not neglected in the low-mass regime, unlike in the heavy-mass case. However, to extract a smooth, averaged profile suitable for normalization, we performed a radial smearing over a few radial periods. In particular, the smeared density is constructed through a Gaussian-like smoothing function, defined as,
\be
\rho_{\rm sg}(r)\to \frac{\int dr' {r'}^2 \rho_{\rm sg}(r)K(r,r')}{\int dr' {r'}^2 K(r,r') },~~~~K(r,r')=e^{-\frac{(\ln r-\ln r')^2}{2\sigma^2}},\label{eq:smearing}
\ee
where $K(r,r')$ is the Gaussian Kernel with $\sigma$ as the smoothing-length parameter \cite{marsh2015nonlinear}. This averaging extracts the effective mean density, allowing for the consistent matching with the NFW-Spike model at the self-gravity radius $r_{\rm sg}$. The density at $r_{\rm sg}$, calculated using the NFW–Spike model, is divided by the corresponding value from our numerical solution and averaged over with the above formula, to determine the overall normalization factor. Subsequently, this normalization factor is applied to the entire density profile. We have illustrated the normalized density profile for the vector DM in the low mass regime in the right panel of Fig.\ref{fig:rhovec1}. In this figure, we observe a clear difference between the scalar and vector DM density distribution. For low masses, spatial variations of the fields become relevant, possibly accounting for the more smeared-out density profile near the horizon in the vector DM scenario. This outcome is consistent with previously obtained numerical results using an initial solitonic profile followed by an NFW tail\cite{Lopez-Sanchez:2025osk}. It is also worth noting that self-interacting scalar DM models have previously been proposed, in contrast to collisionless, free scalar DM models, to achieve such flattening in the density profile \cite{Kadota:2023wlm}. In this context, our present results for vector DM offer a competitive scenario.


\section{Binary Black hole in the vector DM environment}\label{sec.bbh.forces}
So far, we have discussed the behaviour of the vector DM fields in the exterior of the BH and deduced the density profile, which depends on the length scale set by the BH mass and the DM properties. In the following analysis, we will consider a stellar mass BH of mass $m_2$, orbiting the central large mass BH surrounded by the vector DM medium. Particularly, our focus will be on the regime, $q=m_2/M<<1$, often referred to as extreme mass ratio inspiral (EMRI), which acts as the target model for the LISA observation \cite{amaro2017laser, danzmann1996lisa}. In what follows, we will consider the orbit of this binary system to be quasi-circular, with the adiabatic approximation \cite{Kennefick:1995za}. On the other hand, as the binary system is influenced by several dynamical effects, to be discussed next, the orbital frequency becomes effectively time-dependent. The instantaneous orbital frequency is defined as \cite{Cutler:1994ys},  
\be\label{eq:sourcefreq}
\omega^2_s=\frac{GM}{r^3}.
\ee  
The dynamics of the stellar-mass BH evolving in the BH–DM environment are captured in the evolution of the orbital separation, $r$, which determines the orbital frequency. This frequency, in turn, shapes the phase of the GWs emitted by the EMRI, thereby encoding signatures of the surrounding environment of the BH, which will be our topic of discussion in the rest of the part of the present analysis. {\it It is important to note that in the following analysis, we will use the real unit system, explicitly keeping $G$ and $c$ in our calculation.}

\subsection{Radiation reaction force on the binary source} During the evolution of the binary system, the emitted GW carries away energy and angular momentum with it. Hence, the motion of the source, binary system, will be affected. In the quadrapole approximation, the instantaneous energy emission rate for GW takes the following form \cite{maggiore2008gravitational},
\be\label{gwenergyrate}
\frac{dE}{dt}=\frac{G}{5c^5}\dddot{Q}_{ij}\dddot{Q}_{ij},
\ee
where $Q_{ij}$ represents the mass-quadrupole moment tensor of the binary, and the dots represent the derivative with respect to time. Considering the stellar mass BH as a point object, Newtonian particle, this tensor can be quantified as,
\be
Q_{ij}=m_{\rm R}\left(x_ix_j-\frac{1}{3}\delta_{ij}x^2\right),
\ee
where $m_{\rm R}=Mm_2/(M+m_2)$ stands the reduced mass. With the simplified assumption of circular orbit, the coordinate of the center of mass can be written as, $r=(x(t),y(t),0)$, with 
\be
\bea
x(t)=r\cos\omega_s t,\\
y(t)=r\sin\omega_s t.
\eea
\ee
Utilizing this in \eqref{gwenergyrate}, and with the quasi-circular approximation ($\dot{r}=0, \ddot{r}=0$), we arrive at
\be\label{eq:gwpower}
\frac{dE}{dt}=\frac{32Gm^2_{\rm R}r^4\omega^6_s}{5c^5}.
\ee
In the non-relativistic, or Newtonian regime, the emission rate can be expressed in terms of the backreaction force ($F_{\rm GW}$) as follows:
\be
\frac{dE}{dt}={\bf F_{\rm GW}}\cdot {\bf v}.
\ee
where $v=\omega_s r$ (direction is along the tangent of the circular orbit) denotes the velocity of the centre of mass, which can be approximately considered as that of the stellar mass BH for the EMRI system. Nevertheless, considering the force to be isotropic in nature, in the circular plane, the gravitation backreaction force reads
\be\label{eq:gwforce}
{\bf F}_{\rm GW}=\frac{1}{v^2}\frac{32G^4m^2_{\rm R}M^3}{5c^5r^5}{\bf v}.
\ee
This expression, appearing at the lowest post-Newtonian (PN) order, captures the leading contribution to the inspiral dynamics \cite{Cutler:1994ys}. In what follows, this will suffice for the present discussion. 
\subsection{Dynamical friction}\label{sec.dfforce} 
The deceleration of a massive object due to its motion through a surrounding medium has long been recognized \cite{1943ApJ....97..255C, Ostriker:1998fa, Barausse2007}. In a reference frame fixed to the DM environment around the central BH, the orbiting stellar-mass BH experiences such a dissipative force, often referred to as the dynamical friction force, ${\bf F}_{\rm DF}$, acting opposite to its direction of motion. Importantly, this particular effect has been responsible for indirect detection of DM spikes \cite{Chan:2022gqd}. The computation of ${\bf F}_{\rm DF}$, so far, has been mostly done focusing on the scalar field \cite{Hui:2016ltb, Vicente:2022ivh}. For the present case of vector DM. in the particle regime, we will use Chadrashekhar's formula \cite{1943ApJ....97..255C}, which was derived for a generic massive object moving in such a medium. and is given by \footnote{The relativistic correction has been omitted here to provide a leading-order estimate and to ensure all forces are treated consistently at 0PN. For treatments incorporating 1PN and relativistic corrections, see \cite{Speeney:2022ryg, Traykova:2021dua, Barausse2007}.}
\be\label{df.force}
\mathbf{F}_{\mathrm{DF}}=4 \pi \frac{\rho(r) G^2m^2_{\rm R}\mathbf{v}}{v^{3}}\ln \Lambda.
\ee
Where the Coulomb logarithmic term, $\ln \Lambda=b_{\max}v^2_{\rm typ}/(G m_{\rm R})$, with $b_{\rm max}$ denoting the maximum impact parameter and $v_{\rm typ}$ represents the typical velocity of the inspiralling stellar mass BH. Nonetheless, we use the estimated value of this term in the galactic scale, which happens to be $\ln \Lambda\sim 3$ (for rigorous analysis see \cite{just2011dynamical, Just:2004rv}). 

In the wave DM regime, the above expression \eqref{df.force} is usually modified as derived for a scalar field in Ref.~\cite{Hui:2016ltb}.  For the Proca case, as the formula is yet to be derived, 
for approximate estimation, we effectively use the same formula as the scalar \footnote{Previously, in Ref.~\cite{cao2023signatures}, the effect of the dynamical friction force due to higher-spin DM (formed via superradiance) on the binary inspiral has been studied with similar consideration.}, which is given by \cite{Hui:2016ltb} as 
\be
\mathbf{F}_{\mathrm{DF}}=4 \pi \frac{\rho(r) m_{\rm R}^{2} \mathbf{v}}{v^{3}}\left({\rm cin}(2\mu vr)-1+\frac{\sin(2\mu vr)}{2\mu vr}\right),
\ee
where ${\rm cin}(x)=\int^x_0(1-\cos z)dz/z$. Such approximations, treating a vector field through the scalar mode analyses have long been used, particularly in studies of the dynamical Casimir effect \cite{liberati2000sonoluminescence} and radiation from moving mirrors \cite{fulling1976radiation}. However, one should recognize that, although the same recipe is applied to the vector and scalar dynamical forces, differences in their respective density distributions will influence the final results. A detailed, more robust treatment that fully addresses these caveats is left for future investigation.
\begin{figure}
    \centering
\includegraphics[scale=0.38]{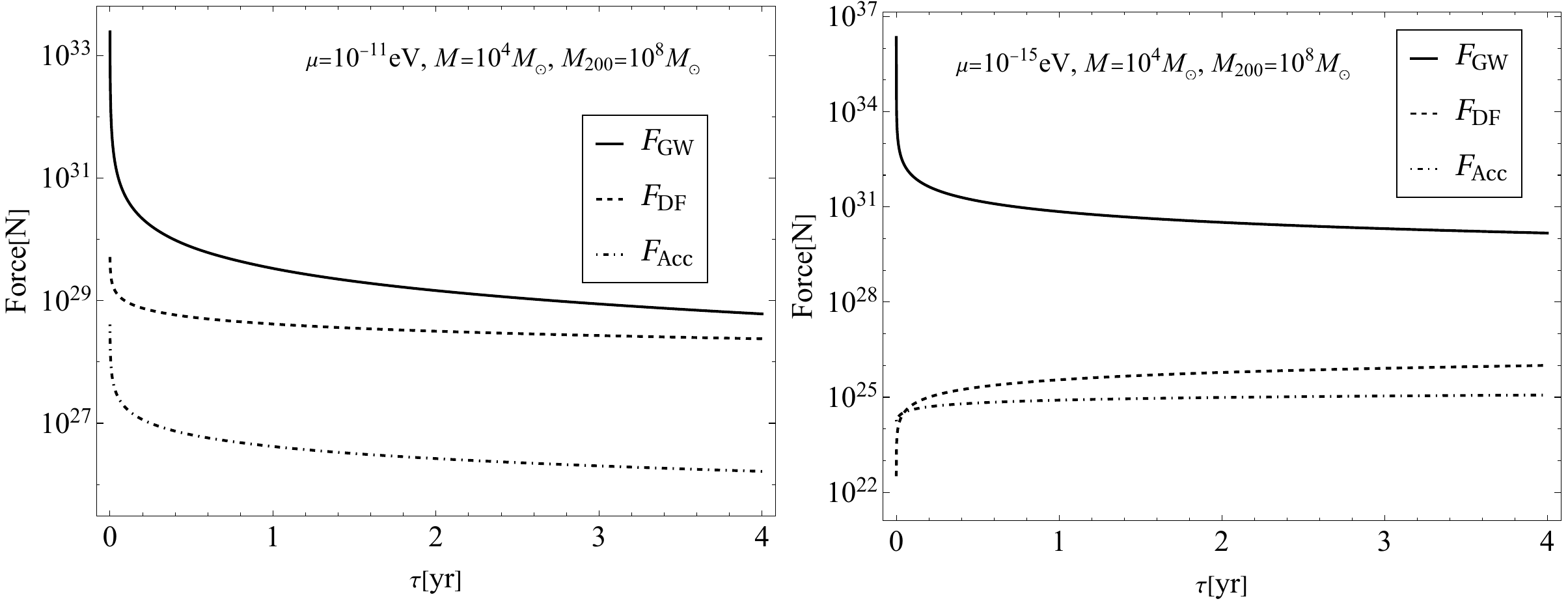}
\caption{Evolution of magnitude of gravitational wave radiation reaction force $F_{\rm GW}(\tau)$  (solid line), dynamical friction force $F_{\rm DF}(\tau)$ (dashed), and the force due to accretion drag $F_{\rm Acc}(\tau)$ (dot dashed line) acting on a stellar mass BH $m_2 = 1 M_{\odot}$ inspiraling into an intermediate-mass black hole of mass $M_{\rm BH} = 10^4 M_{\odot}$. The forces are plotted as a function of $\tau = t_{r = \rm ISCO} - t$.}\label{fig:force}
\end{figure}
\subsection{Accretion of Vector dark matter and drag force on the binary}\label{sec.accretion}
As a natural process, a massive object attracts surrounding matter through its gravitational field. In a binary BH system, this effect can be characterized in terms of the reduced mass, which may have significant implications for the evolution toward the merger phase. In particular, the radial evolution of an EMRI system can be influenced by the drag force induced by matter accretion, which can be quantified as
\be
{\v F}_{\rm Acc}=\frac{dm_{\rm R}}{dt}{\v v}\simeq \frac{dm_2}{dt}{\v v}.
\ee
For a collisionless DM medium in the present scenario, the accretion rate is taken to follow the Bondi–Hoyle–Lyttleton form \cite{bondi1952spherically}, which can be expressed as 
\be
\frac{dm_{\rm R}}{dt}=\frac{16\pi(Gm_2)^2\rho(r)}{vc^2}.
\ee
Nevertheless, the effect of the above force will be studied by solving the radial equation for the EMRI system in the following discussion. 

Combining the three forces, arising due to radiation reaction, dynamical friction and accretion, the evolution of the relative distance between the two BHs can be expressed as
\be
\dot{r}=-(F_{\rm GW}+F_{\rm DF}+F_{\rm acc})\left[2m_{\rm R}\omega_s+m_{\rm R} r\frac{d\omega_s}{dr}\right]^{-1}.
\ee
We find out this trajectory by numerically integrating forward in time, $\tau = t_{r=\rm ISCO} - t$, starting from an initial radius, which is chosen such that the stellar mass binary reaches the innermost stable circular orbit (ISCO) of the central BH, defined by $r=3r_h$, in approximately four years (in time $t=t_{r=\rm ISCO}$). 
In Fig.\ref{fig:force}, the behaviour of the three major forces acting on the inspiral has been presented as a function of $\tau = t_{r=\rm ISCO} - t$ for the particle and wave regimes, in the left and right panels, respectively. As expected, the backreaction due to GW emission dominates the orbital motion, with a force several orders of magnitude larger than the others, both for particle and wave-like DM. The DF force emerges as the next most important effect. Particularly in the particle regime, it surpasses accretion drag by several orders of magnitude. Whereas in the wave regime, the DF force decreases, especially in the near region of the BH, as it is proportional to the density, which is flattened near the core region. In a collisionless DM medium, the accretion drag on an inspiralling object is negligible compared to other dynamical effects, which as illustrated in the Fig.\ref{fig:force}.


\begin{figure}
    \centering    
    \includegraphics[scale=0.5]{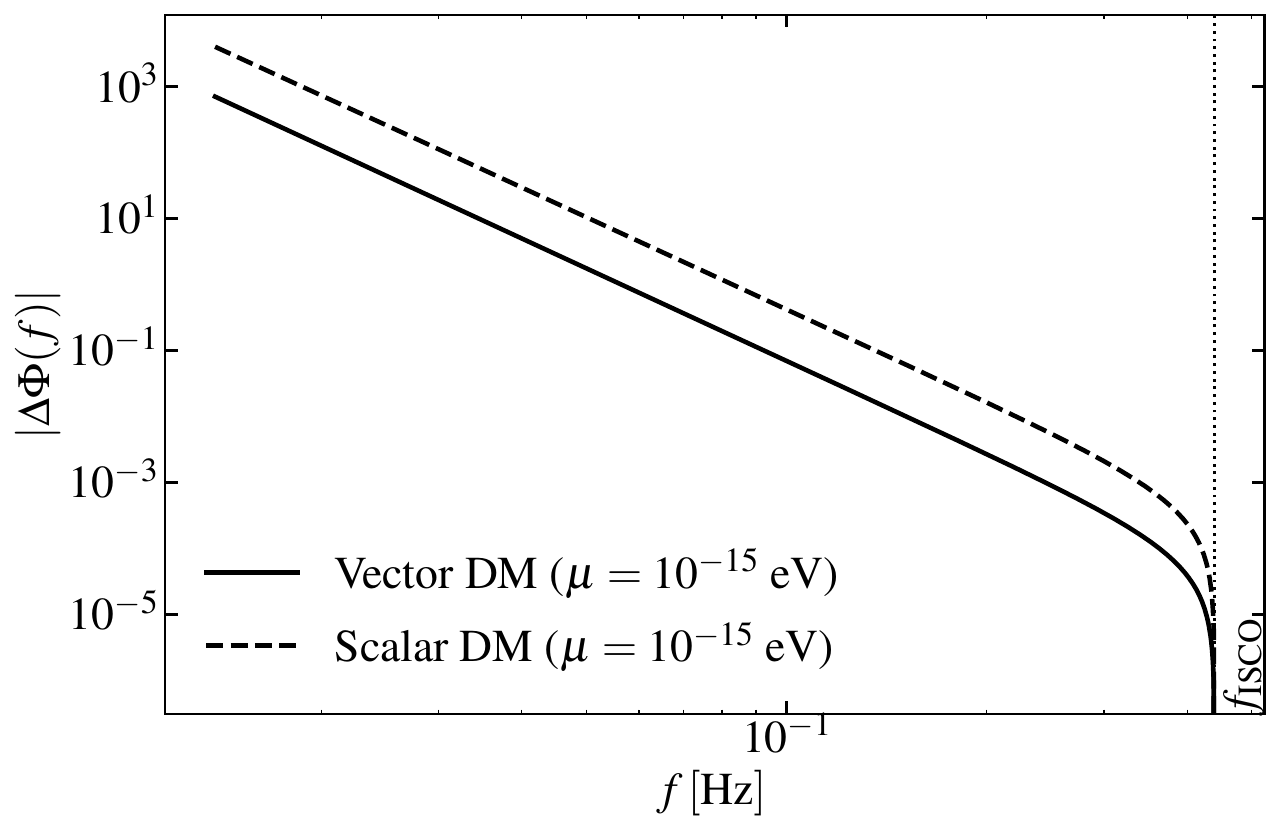}
    \caption{GW dephasing induced by scalar (dashed) and vector DM (dot dashed) with $\mu =10^{-15}\rm eV$. The vertical dotted line stands for the maximum frequency reached up to the ISCO orbit.}
    \label{fig:Dephasing Scalar vs vector}
\end{figure}
\section{Dephasing of gravitational wave due to the presence of vector dark matter}\label{sec.dephasing}
In the preceding discussion, we examined how the surrounding vector DM influences the inspiral through gravitational backreaction, dynamical friction, and accretion. We now turn to how these effects modify the GW signal emitted during the inspiral. The GW signal, as received by the detector, is usually expressed in Fourier space as,
\begin{equation}
    \tilde{h}(f) = F_{+}(\theta,\phi, \chi,f) \tilde{h}_{+}(f) +  F_{\times}(\theta,\phi, \chi,f) \tilde{h}_{\times}(f).
\end{equation}
where $F_{+, \times}$ represent the detector pattern functions corresponding to the polarization states, $\tilde{h}_{+, \times}(f)$. The pattern functions are determined by the polarization angle $\chi$ of the source and by the binary’s position relative to Earth, described by its latitude $\theta$ and longitude $\phi$ in polar coordinates. For simplicity, we consider the strain averaged over these angles, specifically, performing $\langle \tilde{h}(f) \tilde{h}^*(f)\rangle$ \cite{Robson:2018ifk}. In doing so, the expression for the strain reduces to \cite{kim2023adiabatically}, 
\begin{equation}
\tilde h(f) = \sqrt{\frac{4}{5}} A(f)  e^{i\Psi(f)},
\label{eq:htilde}
\end{equation}
with the amplitude,
\begin{equation}
A(f) = \frac{2}{D_L}\left(\frac{G M_c}{c^2}\right)^{5/3}
\left(\frac{\pi f}{c}\right)^{2/3}
\frac{1}{\sqrt{\dot f}},
\label{eq:amplitude}
\end{equation}
where $M_c = (M m_2)^{3/5}/(M + m_2)^{1/5}$ represents the chirp mass of the binary, and $D_L$ stands for the luminosity distance. Whereas, the dot over the frequency denotes the time derivative. Important to note that we restrict our analysis to the Newtonian ($0$PN) approximation, for which one obtain the above amplitude for GW with the stationary phase approximation (SPA) \cite{kim2023adiabatically}. With this approximation, the phase is described as,
\be\label{eq:phase}
\Psi(f) = 2\pi f(t(f)+t_0) - \Phi(f) - \Phi_0 - \frac{\pi}{4},
\ee
with $t_0$ and $\Phi_0$ denoting a reference time and an initial phase, respectively. Whereas $\Phi$ is the phase from the time domain waveform. The  evolution of time and accumulated phase as functions of frequency are given by
\be
t(f) = \int^{f}\frac{df'}{f'},~~~\Phi(f) =2\pi \int^{f}\frac{f'}{\dot{f}'}df',
\ee
with
\be\label{eq.fdot}
\dot f = \frac{P_{\rm GW} + P_{\rm DF} + P_{\rm acc}}{|dE/df|}.
\ee
Where $P_{\rm GW}$ stands for the gravitational-wave power, which can be expressed with the help of \eqref{eq:gwpower} as 
\be
P_{\rm GW}=\frac{dE}{dt}= \frac{32}{5}\frac{c^5}{G}
\left(\frac{G M_c \omega_{\rm GW}}{2c^3}\right)^{10/3}, \qquad \omega_{\rm GW}=2\omega_s=2\pi f.
\ee
In deriving the above expression, we have also utilized \eqref{eq:sourcefreq}. The power lost due to dynamical friction has been represented by $P_{\rm DF}$, computed as $P_{\rm DF} = v F_{\rm DF}$, with $F_{\rm DF}$ denoting the corresponding force as discussed in Sec.\ref{sec.dfforce}. Whereas $P_{\rm acc}$ stands for the power lost due to accretion drag, which can be quantified as $P_{\rm acc} = v F_{\rm acc}$, with $F_{\rm acc}$ representing the effective drag force due to the DM accretion as discussed in Sec.\ref{sec.accretion}. The orbital binding energy, $E(f)$, is modeled in the Newtonian limit as
\be
E(f) = - \left(\frac{G^2 M_c^5 \omega^2_{\rm GW}}{32}\right)^{1/3} .
\ee
With this setup, the dephasing of the GW can be expressed as
\be
\Delta\Phi(f)  = 2\pi \int_{f}^{f_{\text{\rm max}}}df' \left[\left(\frac{f'}{\dot{f}'}\right)_{\text{Vacuum}} - \left(\frac{f'}{\dot{f}'}\right)_{\text{DM Halo}}\right].
\ee
Substituting the expression for $\dot{f}$ from \eqref{eq.fdot}, we have numerically performed the above integration. Importantly, under the circular-orbit approximation, the binary’s separation can be reduced only down to the innermost stable circular orbit (ISCO), which in turn sets the maximum frequency attainable in this approximation (see Ref.\cite{maggiore2008gravitational}). Hence, in the above expression for dephasing, the maximum frequency, $f_{\rm max}$, is set as Min$(f_{\rm high}, f_{\rm ISCO})$, with $f_{\rm high} = 1\rm Hz$ denoting the upper cutoff frequency for LISA \cite{Robson:2018ifk}. Whereas, the minimum frequency, as will be denoted later by $f_{\rm min}$, is set to be $\text{Max}(f_{\rm low}, f_{4\rm yrs})$, with $f_{\rm low} = 0.1 \rm mHz$ representing the lower cutoff frequency for LISA. Within the frequency interval, (${f_{\rm min}}, f_{\rm max}$), we have computed the dephasing in GW, between the vacuum inspiral and the inspiral within the vector DM environment, and presented the results in Fig.\ref{fig:Dephasing Scalar vs vector}. In this figure, we have also provided the dephasing curve for inspirals in the scalar DM case. As the vector fields exhibit a more flattened density profile as compared to the scalar case for low DM mass (see the left panel of Fig.\ref{fig:rhovec1}), they produce distinctive dephasing signatures. The vector case exhibits a milder and more slowly accumulating phase shift as compared to the scalar case. These characteristic differences in the accumulated GW phase indicate that both the magnitude and the frequency dependence of $\Delta \Phi$ are sensitive to the core density profile. With LISA expected to detect low-frequency GW signals in the coming years, it will also be capable of tracking the GW phase evolution, which, according to our analysis, could enable distinguishing between scalar and vector DM via the accumulated phase.

\section{Fisher analysis with LISA detector}\label{sec.fisher}
\begin{figure}
    \centering
    \includegraphics[width=\linewidth]{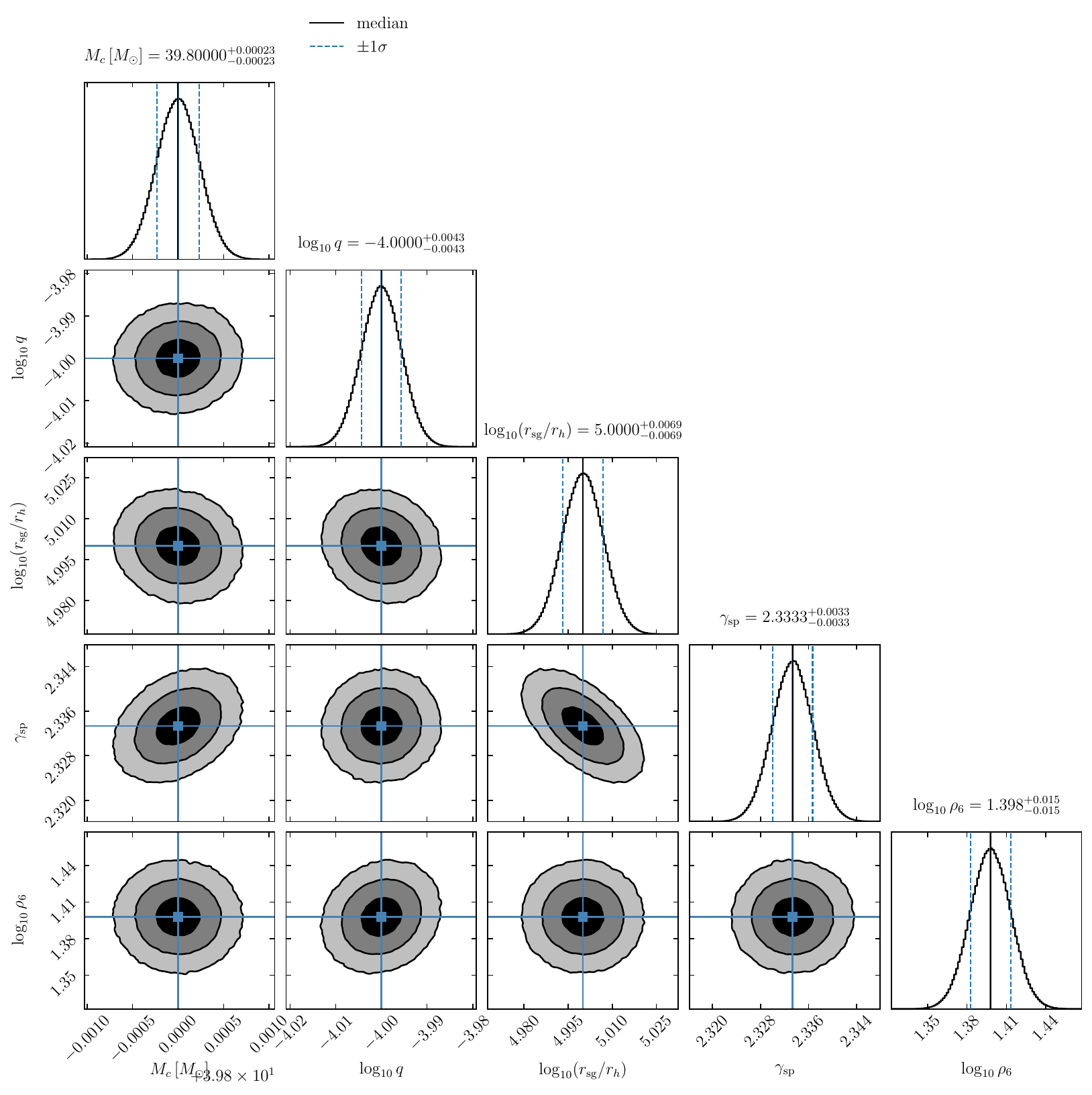}
    \caption{Probability distribution of $\{ M_c[M_{\odot}], \log_{10}q, \log_{10}(r_{\text{sg}/ r_{h}}), \gamma_{\text{sp}}, \log_{10}(\rho_6) \}$ for four years of observation of LISA and $\mu = 10^{-14}\rm eV$. The signal-to-noise ratio is $\rm S/N \simeq14$. The contours indicate the 1-,2-, and 3-$\sigma$ confidence regions. }
    \label{fig: Corner Plot WaveVDM 5e-15}
\end{figure}

\begin{figure}
    \centering
    \includegraphics[width=\linewidth]{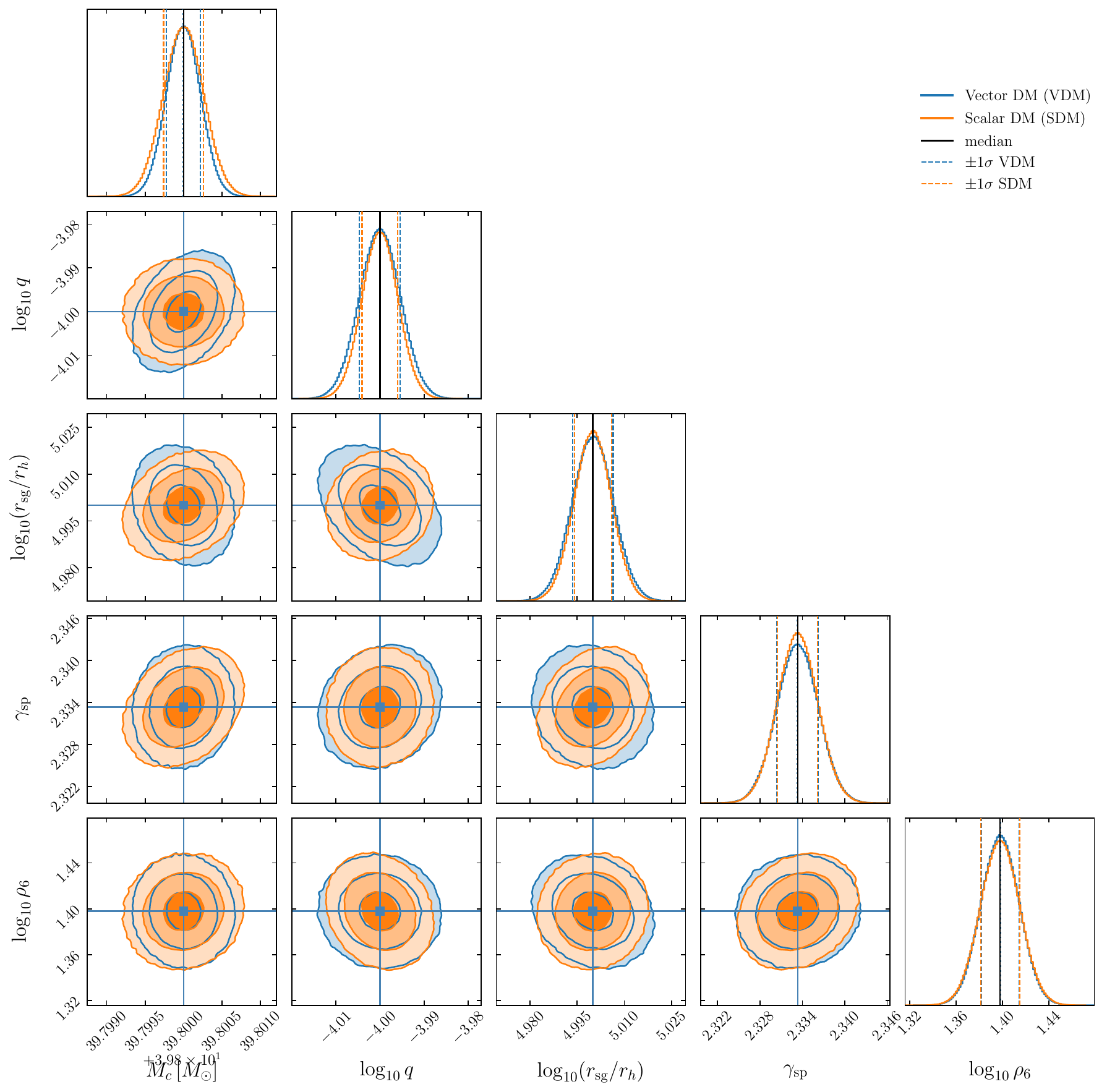}
    \caption{ Probability distributions of $\{ M_c[M_{\odot}], \log_{10}q, \log_{10}(r_{\text{sg}/ r_{h}}), \gamma_{\text{sp}}, \log_{10}(\rho_6) \}$ for four years of observation of LISA for vector and scalar DM of same mass $\mu = 10^{-15}\rm eV$. The signal-to-noise ratio is $\rm S/N \simeq14$. The contours indicate the 1-,2-, and 3-$\sigma$ confidence regions.}
    \label{fig: Corner Plot WaveSDM vs VDM 5e-15ev}
\end{figure}

In the previous section, we discussed the behaviour of the GW phase in a DM environment and showed that, whether treated as particles or waves, and in both scalar and vector cases, the phase difference with respect to the vacuum deviates significantly. Given the high sensitivity of GW on the phase evolution, this will have important implications for identifying the nature of the DM. As LISA will be capable of detecting subtle deviations in the GW phase evolution, the resulting measurements can be used to place constraints on the DM characteristics and its density distribution, along with the BH parameters. Therefore, we perform the Fisher analysis to forecast the values ($\hat{\theta}$) of the source parameters ($\theta$) by analyzing the Fisher information matrix, which reads as \cite{Cutler:1994ys, Barack:2003fp},
\be
\Gamma_{ij}=\left(\frac{\pr h(f)}{\pr \theta_i},\frac{\pr h(f)}{\pr \theta_j}\right)_{\theta=\hat{\theta}},
\ee
with the bracket operation defined as 
\be
(A,B) = 2\int_{f_{\rm min}}^{f_{\rm max}} \frac{A^*(f)B(f) + A(f)B^*(f)}{S_n(f)}df,
\label{Inner Product}
\ee
where $f_{\rm min} = \text{Max}(f_{\rm low}, f_{4 yrs})$ and $f_{\rm max} = \text{Min}(f_{\rm high}, f_{\rm ISCO})$ with $f_{\rm low} = 0.1 \rm mHz$ and $f_{\rm high} = 1 \rm Hz$ are the lower and upper cuttoff frequencies for the LISA and the function $S_n(f)$ denote the noise power spectrum of the detector LISA. In the following analysis, we utilize the analytical fit for $S_n(f)$, as has been obtained in \cite{Robson:2018ifk}, 
\be
S_n(f)=\frac{10}{3L^2}\left(P_{\rm OMS}(f)+\frac{4P_{\rm acc}(f)}{(2\pi f)^4}\right)\left(1+\frac{6}{10}\left(\frac{f}{f_*}\right)^2\right)+S_c(f),
\ee
\begin{figure}
    \centering
    \includegraphics[width=\linewidth]{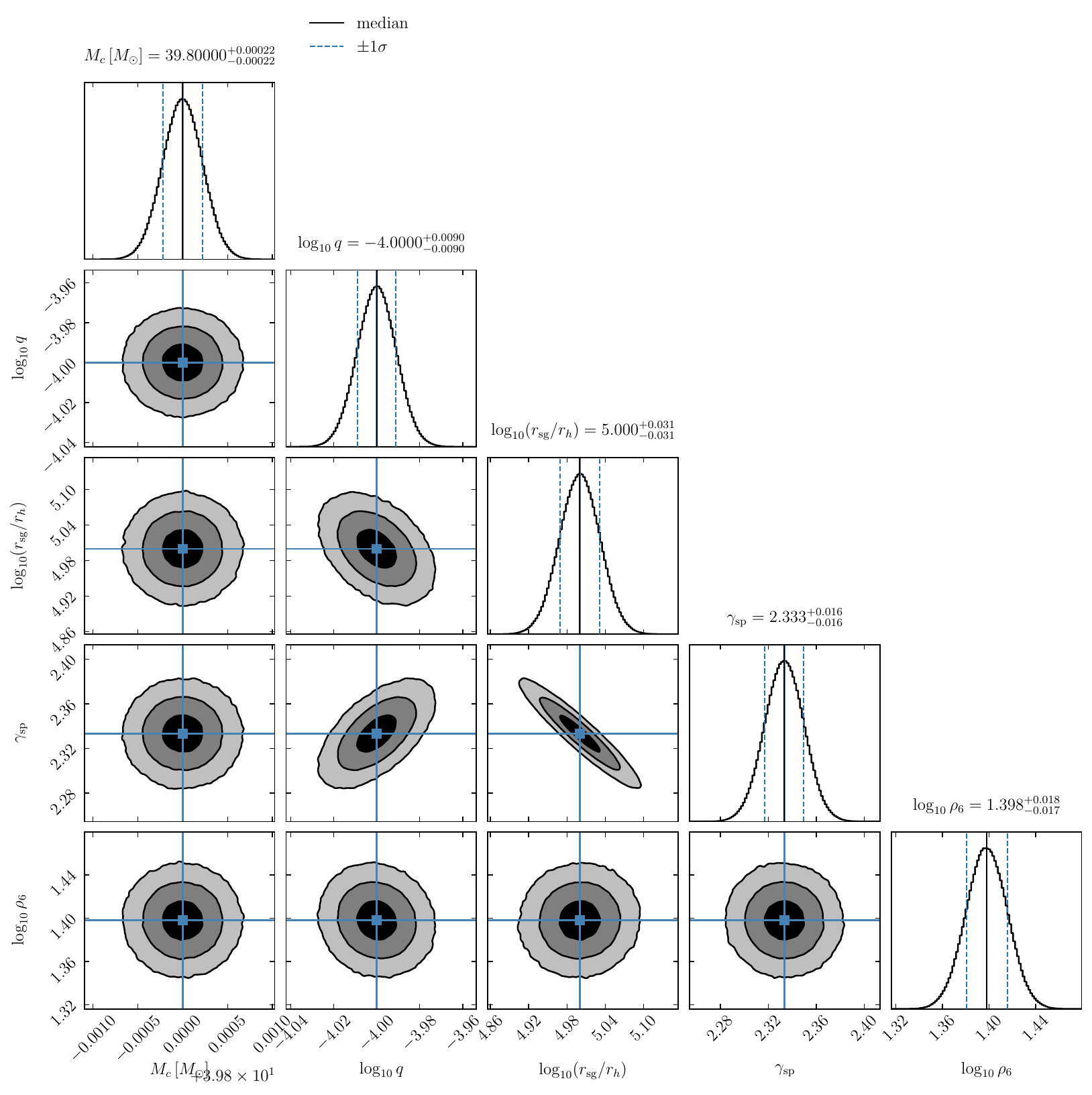}
    \caption{Probability distribution of $\{ M_c[M_{\odot}], \log_{10}q, \log_{10}(r_{\text{sg}/ r_{h}}), \gamma_{\text{sp}}, \log_{10}(\rho_6) \}$ for four years of observation of LISA, with a fixed DM mass, $\mu =10^{-11}\rm eV$ (vector DM in particle regime). The signal-to-noise ratio is $\rm S/N \simeq14$. The contours indicate the 1-,2-, and 3-$\sigma$ confidence regions. }
    \label{fig: Corner Plot Particle VDM}
\end{figure}
with the component noise functions given as
\be
\bea
P_{\rm OMS}(f)&=(1.5\times 10^{-11}{\rm m})^2\left(1+\left(\frac{2 ~{\rm mHz}}{f}\right)^4\right) {\rm Hz}^{-1},\\
P_{\rm acc}(f)&=(3\times 10^{-15}{\rm m~s^{-2}})^2\left(1+\left(\frac{0.4~{\rm mHz}}{f}\right)^2\right)\left(1+\left(\frac{f}{8~{\rm mHz}}\right)^4\right) {\rm Hz}^{-1},\\
S_c(f)&=Af^{-7/3}e^{f^\alpha+\beta f\sin(\kappa f)}\left[1+\tanh{(\gamma(f_k-f))}\right]{\rm Hz}^{-1},
\eea
\ee
where the arm length of the LISA detector is designed to be $L=2.5~{\rm Gm}$. The associated characteristic frequency, $f_*=c/(2\pi L)=19.09~{\rm mHz}$. 
Notably, we have taken into account the optical metrology noise, $P_{\rm OMS}$, acceleration noise due to the test mass, $P_{\rm acc}$, and the confusion noise, $S_c(f)$. The values of the fitting parameters appearing in the expression of $S_c(f)$ are  $\alpha = 0.138, \beta = -221, \kappa = 521, \gamma = 1680, f_{k} = 0.00113$, which have been obtained from Ref.\cite{Robson:2018ifk} for the 4-year observation run of LISA. Given the noise profile, the total signal-to-noise ratio can be measured by computing \cite{Barack:2003fp},
\be
\frac{S}{N} = \sqrt{(h,h)},
\ee
with the inner product defined in \eqref{Inner Product}. Whereas, the root mean squared error, or $1\sigma$ uncertainty, for the individual parameter $\hat{\theta}$ can be expressed as, 
\be
\Delta \theta \equiv \sqrt{\langle (\theta_i-\langle \theta_i\rangle)^2\rangle}=\sqrt{\Sigma_{ii}},
\ee
with $\Sigma_{ii}$ representing the diagonal elements of the inverse Fisher information matrix, $\mathbf{\Sigma} = \mathbf{\Gamma}^{-1}$, commonly referred to as the covariance matrix. The remaining off-diagonal components of the covariance matrix provide the correlation coefficients, 
\be
c_{\theta_i, \theta_j}=\frac{\Sigma_{ij}}{\Delta \theta_i \Delta \theta_j}.
\ee
For the Fisher analysis,  we implement the following set of key intrinsic parameters characterizing the EMRI–vector DM system :
\be
    \theta =\{ M_c[M_{\odot}],  \log_{10}q, \log_{10}(r_{\rm sg}/r_{\rm h}), \gamma_{\text{sp}},\log_{10}\rho_6 \},
\label{eq: Parameters}
\ee
where $\rho_{6}$ is defined as the spike normalization evaluated at $r_6 = 10^{-6}\rm pc$, $\rho_6 = \rho_{\text{sp}}(r_{\text{sp}} / r_6)^{\gamma_{\text{sp}}}$ \cite{Coogan:2021uqv}.
Figure \ref{fig: Corner Plot Particle VDM} shows the corner plot for the probability distribution of the intrinsic parameters $\theta$ for the 4-year LISA observation in the particle regime for vector DM, while Figure \ref{fig: Corner Plot WaveVDM 5e-15} is for the intermidiate regime $(\mu =10^{-14}\rm eV)$ of vector DM where it exibits the particle and wave behavior. In Fig.~\ref{fig: Corner Plot WaveSDM vs VDM 5e-15ev}, we illustrate the comparison of vector and scalar DM for the lowest mass range that we have considered in our analysis, $\mu=10^{-15}{\rm eV}$. For all the three cases we considered $(M_c,q,r_{\rm sg},\gamma_{\rm sp},\rho_6)
=(39.8M_\odot,10^{-4},10^5 r_{\rm h},7/3,25)$ as the set of benchmark values.

The results from the Fisher analysis illustrate LISA's potential to constrain the EMRI parameters in the presence of the vector DM. Across all regimes, the intrinsic parameters, chirp mass $M_{c}$, mass ratio $q$ (these two are determined with high precision) and $\rho_6$ having the same relative uncertainties, $|\Delta \theta_i/\theta_i|$, on the order of $10^{-6}$ for $M_{c}$, $10^{-3}$ for $\log_{10}q$ and $10^{-2}$ for $\log_{10}\rho_6$ within $1\sigma$ (see Table \ref{tab:fisher_mu_scan} and Table \ref{tab:fisher_mu_6e-16}). On the other hand, as we move from particle regime to wave regime up to the intermediate mass range ($\mu=10^{-11}{\rm eV}$ to $10^{-14} {\rm eV}$), we find that the uncertainties in $\log_{10}(r_{\rm sg}/r_{\rm h})$ is not changing, whereas the uncertainties in $\gamma_{\rm sp}$ is changing by 10 times. Moreover, given the similarity in the density profiles between scalar and vector DM cases in the heavy mass regime (see left panel of Fig.~\ref{fig:rhovec1}), we find that the uncertainties in the intrinsic parameters are also appear to be similar. Now for the low mass regime of vector DM $(\mu =  10^{-15}\rm eV)$, in Fig.~\ref{fig: Corner Plot WaveSDM vs VDM 5e-15ev},  we find the uncertainties in $r_{\rm sg}$ and $\gamma_{\rm sp}$ are again changed. In this figure, we have also compared the vector with the scalar DM case. Although both models develop flattened inner density profiles in the low mass regime, they produce different dephasing signatures (see Fig.~\ref{fig:Dephasing Scalar vs vector}). This is due to the way density profiles behave near the central BH with differences in the magnitude in scalar and vector cases (see right panel of Fig.~\ref{fig:rhovec1}). Hence, the dissipative forces, radiation reaction, dynamical friction and accretion drag differ between field types. For this reason, the vector scenario exhibits a different relative uncertainties, specifically concerning the parameter, $\gamma_{\rm sp}$, of the spike density model with relative uncertainties $\sim 1.25\%$ for $\gamma_{\rm sp}$ in the vector case versus $\sim 0.12\%$ in the scalar case ( see Table-\ref{tab:fisher_mu_6e-16}). These results show that both the scalar and vector DM can produce similar uncertainties as far as the parameter estimation is concerned. However, as the wave regime of the DM is more favoured due to uniform density profiles, in this regime the waveform is more informative about $\gamma_{\rm sp}$ in the scalar case.

\begin{table}[h!]
\centering
\begin{tabular}{lcccc}
\hline
Parameter & Fiducial & $\mu = 10^{-11}$ eV & $\mu = 10^{-14}$ eV & \\
\hline
$M_c\,[M_\odot]$              & $39.8$ & $2.19\times10^{-4}$ & $2.46\times10^{-4}$  \\
$\log_{10}q$                  & $-4$   & $6.91\times10^{-3}$ & $4.56\times10^{-3}$  \\
$\log_{10}(r_{\rm sg}/r_{\rm h})$ & $5$     & $2.08\times10^{-2}$ & $6.88\times10^{-2}$ \\
$\gamma_{\rm sp}$             & $7/3$  & $1.04\times10^{-2}$ & $3.29\times10^{-3}$  \\
$\log_{10}\rho_6$             & $\log_{10}25$ 
                             & $1.67\times10^{-2}$  
                             & $1.46\times10^{-2}$ \\
\hline
\end{tabular}

\caption{$1\sigma$ Fisher forecast uncertainties for the benchmark parameters 
$(M_c,q,r_{\rm sg},\gamma_{\rm sp},\rho_6)
=(39.8M_\odot,10^{-4},10^5 r_{\rm h},7/3,25)$ with a 4-year LISA observation.}
\label{tab:fisher_mu_scan}
\end{table}
\begin{table}[h!]
\centering
\begin{tabular}{lccc}
\hline
Parameter & Fiducial & (scalar) &  (vector) \\
\hline
$M_c\,[M_\odot]$                      & $39.8$ & $2.73\times10^{-4}$ & $2.21\times10^{-4}$ \\
$\log_{10}q$                          & $-4$   & $5.29\times10^{-3}$ & $4.63\times10^{-3}$ \\
$\log_{10}(r_{\rm sg}/r_{\rm h})$    & $5$    & $5.43\times10^{-3}$ & $6.45\times10^{-3}$ \\
$\gamma_{\rm sp}$                     & $7/3$  & $2.87\times10^{-3}$ & $2.91\times10^{-2}$ \\
$\log_{10}\rho_6$                     & $\log_{10}25$ 
                                      & $1.71\times10^{-2}$  
                                      & $1.64\times10^{-2}$ \\
\hline
\end{tabular}

\caption{$1\sigma$ Fisher forecast uncertainties for the benchmark parameters 
$(M_c,q,r_{\rm sg},\gamma_{\rm sp},\rho_6)
=(39.8M_\odot,10^{-4},10^5 r_{\rm h},7/3,25)$ with a 4-year LISA observation.
Columns show results for the scalar and vector cases at $\mu=10^{-15}$ eV.}
\label{tab:fisher_mu_6e-16}
\end{table}

\section{Conclusion and outlook}\label{sec.conclusion}
It has already been a decade since the first detection of GWs \cite{LIGOScientific:2016aoc}, marking the birth of the field of GW astronomy. In the years that followed, the study of BH physics has evolved in various directions. Recently, significant attention has turned toward the observation of low-frequency GW. In this context, the space-based detector LISA, with a frequency range of $10^{-4}-1{\rm Hz}$ \cite{amaro2017laser}, has been pivotal in advancing the study of supermassive and intermediate-mass BH binaries, as well as other compact objects (e.g., Neutron stars). Among these, EMRIs with a stellar-mass compact object slowly spiralling into a massive BH in a galactic centre stand out as valuable targets. Having extreme mass ratios, they evolve gradually, emitting long-lived GW signals that remain in LISA’s sensitive band for years and encode detailed information about the spacetime geometry of the central BH and its surroundings. In this pursuit, we consider a stellar mass BH ($1M_\odot$) inspiralling into an intermediate mass BH ($10^4M_\odot$) embedded in a DM medium. 

Building on earlier scalar DM studies \cite{Kadota:2023wlm, kim2023adiabatically}, which simplified the analysis while capturing key observational aspects, in the present work, our primary goal has been to identify features that can reveal the spin of the DM. For this purpose, we consider vector DM, described by the Proca field and present the results alongside the scalar DM case to realize the distinguishability. Alternatively, LISA GW observations \cite{amaro2017laser, danzmann1996lisa} provide an additional motivation, as they can place a variety of constraints on ultralight dark matter that may depend on the spin of the DM candidates. Concerning the observations on DM, which so far provide little information beyond density profiles in galaxies and clusters \cite{Green:2017odb, Famaey:2015bba}, it is imperative to explore a higher-spin DM scenario that reproduces a similar density profile. For this purpose, we combined theoretical models with simulated density profiles that closely resemble observations, specifically the NFW model with a central spike. Because of existing issues with the core density profile, we have explicitly constructed the density profile near the central BH by solving the Proca equations and normalized the solution so that it matches the spike-NFW model in the outer region. On the theoretical side, our analysis reveals that the density profile for the vector DM gets flattened near the BH for DM mass, $\mu\sim 10^{-15}{\rm eV}$. In such mass regimes, the DM field, for both scalar and vector cases, behaves as waves near the black hole. Overall, for higher DM mass, especially in the particle regime, the density profile is steeper because of having a higher transmission into the BH.

To study how different properties of DM influence the GWs from the inspiral of a stellar-mass BH into a central massive BH, three key effects have been included: the backreaction from GW emission, the DF force due to the binary’s motion with respect to the DM, and the drag force from DM accretion. As the binary evolves closer, GW emission intensifies, and the corresponding transfer of energy dominates the orbital evolution. Whereas the DM density flattens near the BH, weakening the gravitational drag and making dynamical friction subdominant. The smoothening effect of the density profile is especially pronounced in the wave-like or low-mass regime. Accordingly, the dominance of GW backreaction over the dynamical friction in the wave DM regime becomes more significant. All of these effects, importantly, carry the information of the density profile. On the other hand, the dephasing of the waveform, which is expected to be an important observable in the context of GW detections \cite{Macedo:2013qea, bertone2024dark, Karydas:2024fcn}, quantifies the difference in accumulated phase shift in the presence of DM relative to the vacuum case. We find that, for inspirals in vector DM, the dephasing varies more gradually with frequency and exhibits clearly distinguishable amplitudes compared to scalar DM. This contrast demonstrates that both the magnitude and frequency dependence encode information about the central density profile. As LISA is expected to observe low-frequency GWs in the near future, tracking the phase evolution and the corresponding frequency-domain data could help distinguish between scalar and vector DM scenarios.

The intrinsic parameters of EMRIs can be strongly constrained in the presence of a surrounding DM environment. Across all regimes, the chirp mass and mass ratio are measured with high precision, with relative uncertainties on the order of $10^{-6}$ and $10^{-3}$, respectively. The constraints on DM spike parameters depend strongly on the particle mass and regime: in the particle and intermediate regimes, uncertainties on the self-gravity radius, slope, and density remain at the few-percent level, while in the wave regime, they degrade significantly. We further performed a comparative analysis between the vector and scalar DM scenarios. Our results indicate that both cases yield comparable parameter uncertainties, with the exception of the spike slope parameter $\gamma_{\rm sp}$. For this parameter, we find that the waveform is more informative (or better constrained) for the scalar DM case than in the vector DM case.

Throughout the entire analysis, there are several points that could be further generalized in future work. The most commonly discussed ultralight DM \cite{Jones:2024fpg} spans the mass range $1-10^{-22}{\rm eV}$ \cite{Ferreira:2020fam}. In the present study, we have focused on the mid-mass window ($10^{-10}$–$10^{-15} {\rm eV}$) for the sake of computational simplicity and for building up the discussion. This choice allows the monopole mode to serve as a reasonable approximation for capturing the most important results, such as GW dephasing, parameter uncertainties, and to show the viability of vector DM alongside the scalar DM scenario. When extending the analysis to the lower mass range, we observed a slight dip in the density profile near the ISCO of BH, indicating the need to go beyond the monopole approximation. Given the present analysis, as well as the earlier study on Proca hair \cite{Hancock:2025ois}, where the procedure for obtaining higher modes is discussed, it should not be difficult to carry out a more thorough analysis regarding the comparison with scalar DM. Next, as far as the GW observation is concerned, we have restricted ourselves to the Newtonian approximation. It directly prompts that for higher accuracy in the measurement of the source parameters, one should include the higher order in PN approximation \cite{Cutler:1994ys}. In that sense, the present work serves as a leading-order estimate of the observables, specifically for static black holes. Since spinning black holes are the most astrophysically relevant \cite{EventHorizonTelescope:2019dse}, analyzing the static case naturally points toward this generalization.\\

\noindent
\textbf{Acknowledgments:}
RK thanks the research group of Xian-Hui Ge for the useful discussions held during the weekly meetings at the Physics Department of Shanghai University. KMV would like to thank Chandra Prakash for fruitful discussions.
 \appendix
\section{Derivation of the solution of Proca equation}\label{Heun_derive}
We rewrite the master equation \eqref{rad.chibar.eqn} of the Proca field in an explicit manner,
\be
\left(1-\frac{r_h}{r}\right)^2\frac{d^2\bar{\chi}}{dr^2}+\frac{r_h}{r^2}\left(1-\frac{r_h}{r}\right)\frac{d\bar{\chi}}{dr}+\left[\omega^2-\left(1-\frac{r_h}{r}\right)\left\{\frac{2}{r^2}-\frac{3r_h}{r^3}+\mu^2\right\}\right]\bar{\chi}=0.
\ee
Rearranging the terms, the above equation takes the following form,
\be
\frac{d^2\bar{\chi}}{dr^2}+\frac{r_h}{r(r-r_h)}\frac{d\bar{\chi}}{dr}+\left(1-\frac{r_h}{r}\right)^{-2}\left[\omega^2-\left(1-\frac{r_h}{r}\right)\left\{\frac{2}{r^2}-\frac{3r_h}{r^3}+\mu^2\right\}\right]\bar{\chi}=0.
\ee
It is convenient to rescale the independent variable in the following manner,
\be
    \bar{\chi}(r)= \sqrt{\frac{r}{r-r_h}}R(r).
\ee
After substitution in the previous equation, the governing equation of $R(r)$ reads
\be
    \frac{d^2R}{dr^2}+ \left[\frac{r_h}{r(r-r_h)^2}-\frac{3r^2_h}{4r^2(r-r_h)^2} + \frac{\omega^2 r^{2}}{(r-r_h)^2} -\frac{2}{r(r-r_h)}+\frac{3r_h}{r^{2}(r-r_h)}-\frac{\mu^2 r}{r-r_h}  \right]R=0.
\ee
It is convenient to redefine the independent variable as $x=r-r_h$ so that the above equation becomes, 
\be \label{general.Rx.eqn}
\frac{d^2R}{dx^2} + \left[k^2+\frac{k^2r_h+\omega^2r_h +\frac{3}{2r_h}}{x} + \frac{\omega^2r^2_h +\frac{1}{4}}{x^2}-\frac{3}{2r_h(x+r_h)} -\frac{15}{4(x+r_h)^2}\right]R =0.
\ee
Considering, $z=-x/r_h$, in the above equation, we arrive at,
\be
\frac{d^2R}{dz^2} + \left[k^2r^2_h-\frac{k^2r^2_h +\omega^2r^2_h +\frac{3}{2}}{z} + \frac{\omega^2r_h^2 +\frac{1}{4}}{z^2}+\frac{3}{2(z-1)} -\frac{15}{4(z-1)^2}\right]R =0.
\ee
Let us express the above equation in the following standard form,
\be
    \frac{d^2 R}{dz^2} + \left[ B_1 + \frac{B_2}{z^2} + \frac{B_3}{z} + \frac{B_4}{(z-1)^2} + \frac{B_5}{z-1} \right] Z=0,
\ee
where the coefficients \( B_1, B_2, B_3, B_4, \) and \( B_5 \) are given as follows:
\begin{align}
B_1 &\equiv -\frac{1}{4} \alpha^2, \\
B_2 &\equiv \frac{1}{4}(1 - \beta^2), \\
B_3 &\equiv \frac{1}{2}(1 - 2\eta), \\
B_4 &\equiv \frac{1}{4}(1 - \gamma^2), \\
B_5 &\equiv \frac{1}{2}(-1 + 2\delta + 2\eta),
\end{align}
with the parameters,
\begin{align}
    \alpha &= \pm 2ikr_h,\\
    \beta &= \pm 2i\omega r_h,\\
    \gamma &= 4,\\
    \delta &= -k^2r^2_h -\omega^2r^2_h, \\
    \eta &= k^2r^2_h + \omega^2r^2_h + 2.
\end{align}
Note that some of the Greek indices are reused in the expressions above. They should not be confused with their earlier appearances in the main text, and their meaning here is restricted to this context. Nevertheless, the above differential equation is the well-known conformal Heun equation \cite{Bezerra:2013iha, Vieira:2021nha}, and the corresponding solution for the above setup can be expressed as
\be \label{append.chibr}
\bea
\bar{\chi}&= c_{1}e^{ikr}r^{3}(r-r_h)^{i\omega r_h}\text{HeunC}\left(-2ikr_h,2i\omega r_h,4,-k^2r^2_h - \omega^2r^2_h,k^2r^2_h+\omega^2r^2_h+2,1-\frac{r}{r_h}\right) \\
&+c_{2}e^{-ikr}r^{3}(r-r_h)^{-i\omega r_h}\text{HeunC}\left(2ikr_h,-2i\omega r_h,4,-k^2r^2_h - \omega^2r^2_h,k^2r^2_h+\omega^2r^2_h+2,1-\frac{r}{r_h}\right).
\eea
\ee
\section{Asymptotic form of the solution}\label{append.asymp.soln}
For large $r$, equivalently, large $x$, the asymptotic form of Eq.~\eqref{general.Rx.eqn} becomes
\begin{equation}
    \frac{d^{2}R}{dx^{2}} + \left[k^2 +\frac{(k^2 + \omega^2)r_h}{x} + \frac{\omega^2r^2_h-\frac{7}{2}}{x^2} \right]R=0.
\end{equation} 
The general solution of this equation allows for the analytical solution,
\be
R(r)\approx a_1M_{-ir_h\left(k+\frac{\mu^{2}}{2k}\right),-\frac{\Omega}{2}}(2ik(r-r_h)) +a_2W_{-i\left(k+\frac{\mu^{2}}{2k}\right),-\frac{\Omega}{2}}(2ik(r-r_h)),
\ee
with $\Omega =i \sqrt{4\omega^{2}-15}$, where $a_1$ and $a_2$ are unknown constant coefficients. The Whittaker's functions, appearing in the above equation, can be reparametrized in terms of the confluent hypergeometric functions as \cite{abramowitz1964handbook},
\be
\bea
&M_{-ir_h\left(k+\frac{\mu^{2}}{2k}\right),-\frac{\Omega}{2}}(2ik(r-r_h))\\
&\equiv e^{-ikr}(r-r_h)^{-\frac{\Omega}{2} + \frac{1}{2}}M\left(ikr_h+\frac{i\mu^2r_h}{2k} + \frac{1-\Omega}{2},-\Omega + 1,2ik(r-r_h)\right),
\eea
\ee
and
\be
\bea
&W_{-ir_h\left(k+\frac{\mu^{2}}{2k}\right),-\frac{\Omega}{2}}(2ik(r-1)) \\
&\equiv e^{-ikr}(r-1)^{-\frac{\Omega}{2} + \frac{1}{2}}U\left(ikr_h+\frac{i\mu^2r_h}{2k} + \frac{1-\Omega}{2},-\Omega + 1,2ik(r-r_h)\right).
\eea
\ee
Moreover, the two hypergeometric functions, $M$ and $U$ are related through the following identity,
\begin{dmath}
U(a,b,z) = \Gamma(b)\Gamma(1-b)\left[\frac{M(a,b,z)}{\Gamma(1+a-b)\Gamma(b)} -z^{1-b}\frac{M(1+a-b,2-b,z)}{\Gamma(a)\Gamma(2-b)}\right].
\end{dmath}
Utilizing this identity, the Whittaker's functions of the second kind can be rewritten as
\be
\bea
W_{-i\left(k+\frac{\mu^{2}}{2k}\right),-\frac{\Omega}{2}}(2ik(r-1)) 
&\equiv e^{-ikr}(r-1)^{-\frac{\Omega}{2} + \frac{1}{2}}\Big[\frac{\Gamma(\Omega)M(ik + \frac{i\mu^2}{2k} + \frac{1-\Omega}{2},1-\Omega,2ik(r-1))}{\Gamma(ik + \frac{i\mu^2}{2k}+\frac{1+\Omega}{2})}\\
&-(2ik(r-1))^{\Omega}\frac{M(ik + \frac{i\mu^2}{2k} + \frac{1+\Omega}{2},1+\Omega,2ik(r-1))}{\Gamma(ik + \frac{i\mu^2}{2k} + \frac{1-\Omega}{2})}\Gamma(1-\Omega)\Gamma(\Omega)\Big].
\eea
\ee
Therefore, it is possible to express the radial solution in terms of a single Hypergeometric function, 
\be
\bea
R(r) &\approx e^{-ikr}(r-1)^{\frac{1-\Omega}{2}}M\left(ik + \frac{i\mu^2}{2k} + \frac{1-\Omega}{2},-\Omega + 1,2ik(r-1)\right) \left[a_1 + \frac{1}{\Gamma(ik + \frac{i\mu^2}{2k}+\frac{1+\Omega}{2})}\right] \\
&+ e^{-ikr}(r-1)^{\frac{1+\Omega}{2}}M\left(ik + \frac{i\mu^2}{2k} + \frac{1+\Omega}{2},\Omega + 1,2ik(r-1)\right) \frac{a_2\Gamma(-\Omega)}{\Gamma\left(ik + \frac{i\mu^2}{2k} + \frac{1-\Omega}{2}\right)}.
\eea
\ee
Finally, we find the radial part of the Proca field in the asymptotic ($r\to \infty$) limit, 
\be\label{append.chi.asymp}
\bea
\chi(r) &\approx a_{3} e^{-ikr}\sqrt{r}(r-r_h)^{\frac{-\Omega}{2}}M\left(ikr_h+ \frac{i\mu^2r_h}{2k} + \frac{1-\Omega}{2},-\Omega + 1, 2ik(r-r_h)\right) \\
&+a_4 e^{-ikr}\sqrt{r}(r-r_h)^{\frac{\Omega}{2}}M\left(ikr_h + \frac{i\mu^2r_h}{2k} + \frac{1+\Omega}{2},\Omega + 1,2ik(r-r_h)\right) 
\eea
\ee
where,
\be
a_3 \equiv \left[a_{1} + \frac{1}{\Gamma(ik + \frac{i\mu^2}{2k}+\frac{1+\Omega}{2})}\right] ; \hspace{0.3cm} a_4\equiv \frac{a_2\Gamma(-\Omega)}{\Gamma\left(ik + \frac{i\mu^2}{2k} + \frac{1-\Omega}{2}\right)}.
\ee
for $k \rightarrow 0$, using $M(a,b,z) \sim \Gamma(b) e^{z/2}\left(\frac{1}{2}bz - az\right)^{\frac{1-b}{2}} J_{b-1}(\sqrt{2bz - 4az}) $
\be
\chi(r) = \sqrt{r}(\mu^2 r_h)^{\Omega/2}\left[a_3 \Gamma(1-\Omega)J_{-\Omega}(2\mu\sqrt{r_h(r-r_h)}) + a_4\Gamma(1+\Omega)J_{\Omega}(2\mu\sqrt{r_h(r-r_h)})\right]
\ee
In the limit $\mu\sqrt{r_hr}>>1$, expanding the Bessel function, the above expression can be cast in the following form,
\be\label{append.chi.exp}
    \chi(r) = r^{1/4}\left(c_3 e^{2i\mu\sqrt{rr_h}}+c_4 e^{-2i\mu\sqrt{rr_h}} \right)+\mathcal{O}\left(\frac{1}{\mu\sqrt{r_hr}}\right),
\ee
where, 
\begin{align}
    c_3&= \frac{e^{i\pi/4}}{\sqrt{\mu\pi}}\left(a_3 \Gamma(1-\Omega) (\mu^2 r_h)^{\Omega/2} e^{-i\pi\Omega/2} + a_4 \Gamma(1+\Omega)(\mu^2r_h)^{-\Omega/2}e^{i\pi\Omega/2} \right),\\
    c_4 &= \frac{e^{i\pi/4}}{\sqrt{\mu\pi}}\left(a_3 \Gamma(1-\Omega) (\mu^2 r_h)^{\Omega/2} e^{i\pi\Omega/2} + a_4 \Gamma(1+\Omega)(\mu^2r_h)^{-\Omega/2}e^{-i\pi\Omega/2} \right).
\end{align}
\section{Energy-momentum tensor of the Proca field}\label{append.proca.density}
The definition of energy-momentum follows from the variation of the action ($\mathcal{S}$) with respect to the metric tensor, and generically reads, 
\be
T_{\mu\nu}=-\frac{2}{\sqrt{-g}}\frac{\delta\mathcal{S}}{g^{\mu\nu}}.
\ee
For a complex vector field, the energy-momentum tensor can be expressed in the following manner \cite{greiner2013field, Hancock:2025ois},
\be
T_{\alpha\beta} = \frac{1}{2}(F_{\alpha\mu}F^{*}_{\beta\nu} + F^{*}_{\alpha\mu}F_{\beta\nu})g^{\mu\nu} -\frac{1}{4}g_{\alpha\beta}F_{\mu\nu}F^{*\mu\nu} + \frac{1}{2}\mu^2(A_{\alpha}A^{*}_{\beta} +A^{*}_{\alpha}A_{\beta} -g_{\alpha\beta}A_{\mu}A^{*\mu}).
\ee
Substituting the mode decomposition \eqref{decom.em} of the Proca field, and expressing it in terms of the gauge invariant variable for $l=0$, the above energy-momentum tensor takes the following form, 
\be
T_{tt} = \frac{\chi^2}{2r^2}f(r)+\frac{1}{2}\mu^2 A^2_t +\frac{f^2 A_r^2 \mu^2}{2}.
\ee
Notably, the individual component appearing in the second and third terms can also be expressed in terms of the gauge invariant variable. Recall that for $l=0$, $A_t = b^{00}/\sqrt{4\pi}$ and $A_r = h^{00}/\sqrt{4\pi}$ \eqref{decom.em}. Using the equation of motion, \eqref{t.eq} and \eqref{r.eq}, we arrive at,
\be
 b^{00}=-\frac{f}{\mu^2 r^2}\partial_r(r\chi), ~~~ 
h^{00} = -\frac{\partial_t\chi}{f\mu^2 r}.
\ee
Finally, the energy-momentum tensor, fully in terms of the gauge invariant variables, reads,
\be
T_{tt} =\frac{1}{4\pi}\left[\frac{\chi^2 f(r)}{2r^2} +\frac{(\partial_t \chi)^2 }{2\mu^2r^2} +\frac{f^2(r)}{2\mu^2 r^4}\left[\partial_r(r\chi)\right]^2\right].
\ee
Substituting $\chi(t,r) = \bar{\chi}(r)e^{-i\mu t}$, and considering the large-mass limit, so that one can {\it neglect the spatial gradient term}, we get the above expression in the following simplified form,
\be
T_{tt} = \frac{|\bar{\chi}(r)|^2}{4\pi r^2}\left[\frac{1}{2}\left(1+f(r)\right)+\frac{f^2(r)}{2\mu^2 r^2}\right].
\ee
The DM density, therefore, can be evaluated as \cite{Ravanal:2024odh},
\be
\rho=-{T^t}_t=-g^{tt}T_{tt}=\frac{|\bar{\chi}(r)|^2}{4\pi r^2f(r)}\left[\frac{1}{2}\left(1+f(r)\right)+\frac{f^2(r)}{2\mu^2 r^2}\right].
\ee
On the other hand, the radiation flux component, required for computing the incoming and outgoing flux rate (as will be required for the discussion in Appendix \ref{append.transmission}), is given by
\be
{T^r}_t=g^{rr}T_{rt} = \frac{1}{2}\mu^2g^{rr}(A^*_r A_t+ A^*_tA_r).
\ee
Performing a similar manipulation as above, this component reads in terms of the gauge invariant variables as,
\be
{T^r}_t =\frac{g^{rr}}{8\pi \mu^2 r^{3}}[(r\partial_{r}\chi +\chi)\partial_{t}\chi^{*} + (r\partial_{r}\chi +\chi)^{*}\partial_{t}\chi].
\label{Trt}
\ee
\section{Considering the DM to be described by a massive scalar field}\label{append.scalar.density}
In Schwarzschild spacetime, the behaviour of a minimally coupled massive scalar field, has been studied for a wide range of masses in \cite{Hui:2019aqm}. In the heavy DM mass regime, for which $\mu\geq r_h^{-1}$, as considered for vector DM, the solution with ingoing near-horizon condition is given by 
\be
\phi(t,r)\simeq \phi_0 e^{-i\mu t}\frac{1}{r^{3/4}} e^{-2i\mu\sqrt{rr_h}}\left(1+\mathcal{O}\left(\frac{1}{\mu\sqrt{r}}\right)\right),
\ee
where $\phi_0$ stands for the overall constant amplitude to be fixed through normalization utilizing the density profile. Whereas, $\mu$ denotes the mass of the scalar field. Substituting this into the energy-momentum tensor for the scalar field (neglecting the radial gradient), we arrive at
\be
\rho=-{T^t}_t=-g^{tt}T_{tt}=\frac{1}{4\pi f(r)}\mu^2\phi^2(t,r)=\frac{1}{4\pi f(r) r^{3/2}}\mu^2\phi^2_0.
\ee
Utilizing the NFW profile with the spike model as discussed in Sec.\ref{sec.model.density}, the normalized solution, up to the leading order, reads 
\be
\phi(t,r)\simeq \sqrt{4\pi f(r_{\rm sg})\rho_{\rm sp}\left(\frac{r_{\rm sg}}{r_{\rm sp}}\right)^{-\gamma}}\frac{r_{\rm sg}^{3/4}}{r^{3/4}\mu}e^{-i\mu t}e^{-2i\mu\sqrt{r r_h}}.
\ee
Substituting this normalized solution again in the energy-momentum tensor for the scalar field, the density profile can be extracted as 
\be
\rho=\frac{1}{2f(r)}\left[|\pr_t\phi|^2+f^2(r)|\pr_r\phi|^2+\mu^2f\phi^2\right]=f(r_{\rm sg})\rho_{\rm sp}\left(\frac{r_{\rm sg}}{r_{\rm sp}}\right)^{-\gamma}\frac{r_{\rm sg}^{3/2}}{r^{3/2}f(r)}\left[1+f^2(r)\frac{9+16rr_h\mu^2}{32r^2\mu^2}\right].
\ee
Whereas, in the low mass regime, we follow the numerical procedure as depicted in Sec.\ref{sec.field.eqn.soln} and utilize the general solution in terms of the confluent Heun function with ingoing boundary condition near the horizon,
\be
\phi(t,r)=\phi_0e^{-i\omega t}(r-1)^{-i\omega}e^{-ikr}\text{HeunC}(2i\Bar{k},-2i\omega,0,-\omega^{2}-k^2,\omega^{2}+k^{2}-l(l+1),1-r)
\ee
where, as usual, $k=\sqrt{\omega^2-\mu^2}$.
\section{Computation of reflection and transmission coefficients of the scattered vector dark matter (particle regime)}\label{append.transmission}
The procedure of finding the reflection and transmission coefficients for a matter field begins with the identification of the scattering states at a large distance (in principle at spatial infinity) from the BH horizon \cite{unruh1976absorption}. The analysis of field modes in the main text has been restricted within the self-gravity radius $r_{\rm sg}$. It is imperative that one needs to go beyond this region for the asymptotic nature of the scattering states. However, in the region beyond $r_{\rm sg}$, where the self-gravity becomes crucial, the metric should be modified and can be approximately given as \cite{Hui:2019aqm},  
\be
ds^2 = -f(r_{\rm sg}) dt^2 + \frac{dr^2}{f(r_{\rm sg})}+r^2d\theta^2+r^2\sin^2\theta d\varphi^2,
\ee
where $f(r_{\rm sg})=1-r_h/r_{\rm sg}$. As we will see, this consideration leads to the effective potential of the vector DM particles not vanishing at spatial infinity but instead approaching a constant value. In this region of the spacetime, the governing equation of the gauge field reads
\be\label{beyond.rsg}
\partial_{r_*}^{2}\bar{\chi} + \left[k^2 +\frac{\mu^2 r_h}{r_{\rm sg}}-\frac{2}{r_{*}^{2}}\right]\bar{\chi} = 0,
\ee
Notably, here, we should mention the Tortoise coordinate, following the definition $dr_*=dr/f(r_{\rm sg})$.
Using leading order approximation for massive DM fields, and defining $\bar{k}^2 = k^2 + (\mu^2 r_h/r_{\rm sg})$, we express \eqref{beyond.rsg} as,
\be
\partial_{r_*}^{2}\bar{\chi}+\bar{k}^2\bar{\chi} = 0.
\ee
The general solution of this equation can be expressed as,
\be
\bar{\chi}(r)=b_1e^{\mi\bar{k} r_*}+b_2e^{-\mi\bar{k} r_*}. 
\label{Outside rsg}
\ee
Therefore, the solution is essentially the superposition of ingoing and outgoing waves as can be realized by associating the time part,
\be\label{soln.rgtrsg}
\chi(t, r)=e^{-\mi \omega t}\left[b_1e^{\mi\bar{k} r_*}+b_2e^{-\mi\bar{k} r_*}\right].
\ee
Whereas, for $r<r_{\rm sg}$, considering the particle DM  case, with $\mu r_{h}\gg 1$, we have \eqref{soln.chi.heavymass},
\be
    \chi(r) =c_{4} r^{1/4} e^{-2i\mu\sqrt{rr_h}}.
\ee
Imposing the continuity at $r=r_{\rm sg}$ for $\chi(r)$ and $\partial_{r}\chi(r)$ yields
\be
\frac{c_4}{b_2} = r_i^{-1/4}e^{i\mu \sqrt{r_{\rm sg}r_h}} + \mathcal{O}\left(\frac{1}{\mu\sqrt{r_{\rm sg} r_h}}\right),
\ee
and 
\be
\frac{b_1}{b_2}  = \frac{1}{8 i \mu \sqrt{r_{\rm sg}r_h}} e^{-2i\mu\sqrt{r_{\rm sg}r_h}} +\mathcal{O}\left(\frac{1}{(\mu\sqrt{r_{\rm sg} r_h})^2}\right).
\ee
Using \eqref{Outside rsg} in \eqref{Trt} we obtain the 
incident ($\mathcal{E}_{\rm in}$) and reflected parts ($\mathcal{E}_{\rm out}$) of the flux over a spherical surface at a large radial distance $r<r_{\rm sg}$, respectively, as \cite{Cardoso:2019dte}, 
\be
\mathcal{E}_{\rm in}=4\pi r^2\left({T^r}_t\right)_{\rm in} = \sqrt{\frac{r_h}{r_{\rm sg}}}|b_2|^2, \hspace{0.2cm} \mathcal{E}_{\rm out}=4\pi r^2\left({T^r}_t\right)_{\rm out} = \sqrt{\frac{r_h}{r_{\rm sg}}}|b_1|^2.
\ee
Therefore, the reflection coefficient can be extracted as,
\be
\mathcal{R}=\frac{\mathcal{E}_{\rm out}}{\mathcal{E}_{\rm in}} \approx \frac{1}{16 \mu^2 r_{\rm sg} r_{h}} 
\ee
From the above expression, it follows that in the particle regime, $\mu r_h\gg 1$, the reflection coefficient should vanish, $\mathcal{R}\sim 0$. On the other hand, the flux conservation dictates the transmission coefficient $\mathcal{T}$ to be 
\be
\mathcal{T} \sim 1-\mathcal{R} \sim 1.
\ee
\bibliographystyle{JHEP}
\bibliography{Bibliography}
\end{document}